\documentclass[12pt]{article}
\usepackage{epsfig,amssymb,amsmath,amscd}
\usepackage{latexsym}

\newcommand{\bq}{\begin{equation}}
\newcommand{\eq}{\end{equation}}
\newcommand{\bqa}{\begin{eqnarray}}
\newcommand{\eqa}{\end{eqnarray}}
\newcommand{\ben}{\begin{enumerate}}
\newcommand{\een}{\end{enumerate}}
\newcommand{\bc}{\begin{center}}
\newcommand{\ec}{\end{center}}
\newcommand{\bqb}{\begin{eqnarray*}}
\newcommand{\eqb}{\end{eqnarray*}}

\newcommand{\psl}{\rlap / p}

\newcommand{\qpsl}{\rlap / q'}
\newcommand{\ppsl}{\rlap / p'}
\newcommand{\qppsl}{\rlap / q''}

\def\pr#1#2#3{Phys. Rev. ${\bf{#1}}$, #2 (#3)}
\def\prl#1#2#3{Phys. Rev. Lett. ${\bf{#1}}$, #2 (#3)}
\def\pl#1#2#3{Phys. Lett. ${\bf{#1}}$, #2 (#3)}
\def\prep#1#2#3{Phys. Rept. ${\bf{#1}}$, #2 (#3)}

\def\np#1#2#3{Nucl. Phys. ${\bf{#1}}$, #2 (#3)}

\def\jhep#1#2#3{JHEP ${\bf{#1}}$, #2 (#3)}

\def\ijmp#1#2#3{Int. J. Mod. Phys. ${\bf{#1}}$, #2 (#3)}

\def\fortp#1#2#3{Fortsch. Phys. ${\bf{#1}}$, #2 (#3)}

\def\polon#1#2#3{Acta Phys. Polon. ${\bf{#1}}$, #2 (#3)}

\def\jmp#1#2#3{J. Mod. Phys. ${\bf{#1}}$, #2 (#3)}

\begin{document}
\pagenumbering{arabic}
\thispagestyle{empty}
\def\thefootnote{\fnsymbol{footnote}}
\setcounter{footnote}{1}

\begin{flushright}
May 7, 2016\\
 \end{flushright}

\vspace{2cm}

\begin{center}
{\Large {\bf Test of the triple Higgs boson form factor in $\mu^-\mu^+\to HH$}}.\\
 \vspace{1cm}
{\large G.J. Gounaris$^a$ and F.M. Renard$^b$}\\
\vspace{0.2cm}
$^a$Department of Theoretical Physics, Aristotle
University of Thessaloniki,\\
Gr-54124, Thessaloniki, Greece.\\
\vspace{0.2cm}
$^b$Laboratoire Univers et Particules de Montpellier,
UMR 5299\\
Universit\'{e} Montpellier II, Place Eug\`{e}ne Bataillon CC072\\
 F-34095 Montpellier Cedex 5.\\
\end{center}

\vspace*{1.cm}
\begin{center}
{\bf Abstract}
\end{center}

We study the sensitivity of the process $\mu^-\mu^+\to HH$ to the
$q^2$-dependence of the $HHH$ form factor, which can reflect the Higgs boson
structure, especially in the case of compositeness.
We compute the Born and 1-loop SM contribution to this
process. We then show how the $\mu^-\mu^+\to HH$ polarized and unpolarized
cross sections are modified by the presence of various types of
anomalous contributions to the $HHH$ form factor, in particular Higgs constituents
in the case of compositeness.

\vspace{0.5cm}

\def\thefootnote{\arabic{footnote}}
\setcounter{footnote}{0}
\clearpage

\section{Introduction}

In spite of the discovery \cite {Higgsdiscov} of the Higgs boson \cite {Higgs},
as expected in the standard model (SM) \cite{SM},
the SM cannot be the last word, and physics beyond SM (BSM) should exist \cite{revBSM}.
A review on Higgs physics is for example made in \cite{SMHiggsrev}.
In respect of BSM various types of proposals have been made leading for example
to anomalous Higgs boson couplings \cite{Anom1,Anom2}
or to couplings of the Higgs boson
to new visible or invisible particles \cite{Portal},  particularly in the case
of Higgs boson compositeness \cite{Hcomp1, Hcomp2,
partialcomp, Hcomp3, revH}.

There are many processes (involving Higgs boson production or decay)
where such  Higgs boson couplings, differing from SM predictions,
could be observed. However, in most of them the Higgs boson
is on-shell and such a  departure could not obviously tell whether it
is caused by Higgs  compositeness.
This is particularly the case if the Higgs boson is coupled to an invisible
sector.

The observation  of a suitable Higgs boson
form factor though, could give an answer to such questions.
Indeed a composite particle (like the proton or the pion) should
have a  form factor. But contrarily to the case of the proton and
the pion, we do not have a $\gamma HH$ or a $ZHH$ vertex for studying
the $H$ form factor.\\

In this paper we concentrate on  the $HHH$ form factor;
that is at a $q^2\equiv s$ dependence of the $HHH$ vertex when one $H$,
with momentum $q$, is off-shell. The scale of this dependence could
be in the  TeV range, see e.g.  \cite{Hcomp1}.

There are various processes which
involve the $HHH$ coupling, but not necessarily the form factor
with one $H$ far off-shell, in which we are interested.
Such processes at LHC are $g g \to H \to HH$ (with $g$ denoting a gluon) or
$ZZ \to H  \to HH$, $W^+W^- \to H  \to HH$ \cite{HHLHC};
but these require complex theoretical and experimental analyses,
before reaching the structure of the $HHH$ coupling.
Similarly, the $\gamma\gamma\to H  \to HH$ process also involves
a complex initial  1-loop $H$ coupling related to the $HHH$ form factor.

The simplest process directly sensitive to an $HHH$ form factor effect,
is then probably $\mu^-\mu^+\to HH$, observable at a future $\mu^-\mu^+$ collider \cite{mumucol}.
Even if this will be realizable only in a far future and under the condition
of obtention of a very high luminosity, this process is
particularly interesting because in non standard Higgs models, the $H\mu\mu$
coupling may still be similar to the SM one, whereas the Higgs boson
self-coupling and the Higgs boson couplings
to heavy fermions may be very different, see e.g. \cite{partialcomp}.

So below we analyze $\mu^-\mu^+\to HH$  in this spirit.
We start by computing the SM Born and 1-loop contributions to the helicity
amplitudes and cross sections.
At Born level they are due to $s$-channel $H$ exchange and (at a weaker level)
to $t$ and $u$ channel $\mu$ exchange.
At 1-loop level, the corrections involve various triangles, boxes as well as
$\mu$ and $H$ self-energy bubbles. Some of these terms (the  ``right triangles" and
the  ``bubbles with 4-leg couplings and $H$ self-energy diagrams")
already create contributions to the $HHH$ form factor
from the usual scalars, fermions and gauge boson loops, but they are relatively
small, at order $\alpha$.
We collect them in the appendix and we illustrate their corresponding (modest) $s$ dependence.

We then compute examples of new contributions which
could be induced  either by Higgs boson compositeness or by
the couplings of the Higgs boson to a new set of particles.
Illustrations show how such contributions can generate spectacular
differences in the $s$ dependence of the $HHH$ form factor,
with respect to the one predicted by the SM.

These different 1-loop corrections reflect in the various amplitudes
and cross sections and   could be useful
for guessing what type of contribution is necessary in order to explain a possible
departure of the measurements with respect to the SM expectation.
We separately consider the helicity conserving (HC) and the helicity
violating (HV) amplitudes and the polarized cross sections, as these ones
may be measurable in this process \cite{mumupol}. \\

Contents: Section 2 is devoted to the presentation of the
SM Born amplitudes and cross sections of the $\mu^-\mu^+\to HH$ process
and Section 3 to the SM 1-loop contributions. Examples of anomalous HHH contributions
effects are described in Section 4.
In the concluding Section 5 we summarize our results and we mention that
this type of study of the effect of
the $HHH$ form factor could also be done in several other (but more complex)
$HH$ production processes.\\

\section{SM Born amplitudes and cross sections}

The SM Born amplitude of the $\mu^-\mu^+\to HH$ process is due to 3 diagrams:
the $s$ channel $H$ exchange with an initial $(\mu^- \mu^+ H)$ and a final $(HHH)$ coupling,
and two diagrams with $t$ and $u$ channel $\mu^\mp$ exchange with
up and down $(\mu^- \mu^+ H)$ couplings.
The invariant Born amplitude is
\bqa
&&A^{\rm Born}(\mu^-\mu^+\to HH)=-~{e^2g_{\mu\mu H}g_{HHH}\over s-m^2_H+im_H\Gamma_H}\bar v(l',\lambda')
u(l,\lambda)\nonumber\\
&&-e^2g^2_{\mu\mu H}\left [{\bar v(l',\lambda')(\qpsl+m_{\mu}) u(l,\lambda)\over t-m^2_{\mu}}
+{\bar v(l',\lambda')(\qppsl +m_{\mu})u(l,\lambda)\over u-m^2_{\mu}} \right ]~~, \label{ABorn-inv}
\eqa
\noindent
where $(\lambda,\lambda')$ are the $(\mu^-,\mu^+)$ helicities, $(l,l',p,p')$ are the $(\mu^-,\mu^+,H,H)$
momenta and we also define
\bqa
&& q=l+l' ~, ~ q'=l-p=p'-l' ~,~ q''=l-p'=p-l'~, \nonumber \\
&& s=q^2 ~,~  t=q'^2 ~,~ u=q''^2 ~,~ \label{kinem1}
\eqa
and the couplings
\bq
g_{\mu\mu H}=-{m_{\mu}\over2s_Wm_W}=-{m_{\mu}\over e v} ~~, ~~
g_{HHH}=-{3m^2_{H}\over2s_Wm_W}=-{3m^2_{H}\over e v} ~, \label{couplings1}
\eq
where  the final $HH$ are symmetrized.

The corresponding Born helicity amplitudes are
\bqa
&&F^{\rm Born}_{\lambda,\lambda'}(s,\theta)=
{e^2g_{\mu\mu H}g_{HHH}\over s-m^2_H+im_H\Gamma_H}\sqrt{s}\delta_{\lambda,\lambda'}\nonumber\\
&&+e^2g^2_{\mu\mu H}\left [{1\over t-m^2_{\mu}}-{1\over u-m^2_{\mu}}\right ](2\lambda)
p_H\sqrt{s}\sin\theta \delta_{\lambda,-\lambda'} ~~, \label{FBorn-helicity}
\eqa
where $p_H=\sqrt{s/4-m^2_H}$, and $\theta$ is the c.m. scattering angle between $l$ and $p$.
Note that we computed them from the invariant amplitude in (\ref{ABorn-inv}) by neglecting
the $m_{\mu}/\sqrt{s}$ terms  appearing in the $\mu$ propagator and in
the precise expressions of the Dirac spinors. These Born amplitudes
are already factorized by one or two $g_{\mu\mu H}$ couplings proportional to $m_{\mu}/ m_W $,
so that there is no need for keeping these negligible corrections.

Note also that  s-channel part in (\ref{FBorn-helicity}) is angle-independent and
purely helicity violating (HV), due to the $\delta_{\lambda,\lambda'}$ term, which
violates the high energy helicity conservation rule
$\sum \lambda_{in}=\sum \lambda_{fin}$ \cite{heli}. This term dominates
at low energy though, but it decreases like $1/\sqrt{s}$ as the energy increases.

The $t$ and $u$ channel parts are purely helicity conserving (HC), when neglecting
$m_{\mu}/\sqrt{s}$ terms. They tend to a constant at high energy
and are forward-backward antisymmetric (vanishing at $\pi/2$).
They are about 100 times weaker than the $s$-channel part though,  because
of their additional small $g_{\mu\mu H}$ coupling factor.

So finally the only non negligible Born amplitudes are the HV ones
(i.e. those due to the s-channel $H$ exchange) in which
we are interested, because of their proportionality to the $HHH$ coupling.
This is the first remarkable feature of the $\mu^-\mu^+\to HH$ process.\\

The cross section for unpolarized $\mu^{\mp}$ beams is
\bq
{d\sigma\over dcos\theta}={p_H\over 64\pi s\sqrt{s}}\sum_{\lambda,\lambda'}
|F_{\lambda,\lambda'}(s,\theta)|^2 ~~. \label{dsigma-unpol}
\eq
Cross sections with left-handed or right-handed polarized $\mu^\mp$
beams will also be considered.
Note that due to the final $HH$ symmetrization,  the cross sections are necessarily
forward-backward symmetric.\\

\section {SM 1-loop contributions.}

The 1-loop corrections to the above Born terms contain various types of diagrams;
triangle diagrams for initial and final vertices and $H$ self-energy
bubbles for $s$-channel $H$ exchange;  triangle diagrams for up or down
vertices and $\mu^\mp$ self-energy bubble for $t,u$ channel $\mu^\mp$ exchange;
several types (direct, crossed and twisted) of box diagrams. Some of the triangles
and bubbles are divergent and a choice of renormalization scheme has
to be made, consisting in the addition of specific counter terms canceling these
divergences.
There are various schemes for this, which differ by their choice of experimental
inputs, see \cite{OS}. One may for example use the on-shell (OS) scheme;
a special application to SM and MSSM Higgs couplings is done in \cite{HHHOS}.

However in the present study we are essentially interested in the $s$ dependence
of the $HHH$ vertex (to be then compared with possible new physics effects)
and not in its precise renormalized on-shell value which will be difficult
to measure accurately anyway.
For this purpose we will compute the various 1-loop terms in the SRS
scheme \cite{SRS} which give simple high energy expressions, whose contents
are immediately readable and instructive,
 in particular for suggesting possible models for new contributions.\\

We  next list the various diagrams and their relative importance.

\underline{{\bf (a) In the s-channel sector}} we find:\\

{\bf (a1)  Left triangles}
\[
(W\nu W), (Z\mu Z), (\mu Z\mu), (G\nu G), (G^0\mu G^0),
(H\mu H), (\mu G^0\mu), (\mu H\mu)~,
\]
followed by $s$-channel  $H$ exchange and a final $HHH$ coupling.\\

{\bf (a2)  Left triangles connected to the final $HH$ by a 4leg coupling}
\bqa
&& (W\nu W)+(WWHH) ~,~ (Z\mu Z)+(ZZHH) ~,~ (G\nu G)+(GGHH) ~,~ \nonumber \\
&& (G^0\mu G^0)+(G^0G^0HH) ~,~ (H\mu H)+(HHHH) ~. \nonumber
\eqa

Denoting by $T^{\rm SM}_{\rm left}$ the sum of the a1 and a2 diagrams,
the implied helicity amplitude is written as
\bq
F^{\rm left~SM}_{\lambda,\lambda'}=
{eT^{\rm SM}_{\rm left}(s)g_{HHH}\over s-m^2_H+im_H\Gamma_H}\sqrt{s}\delta_{\lambda,\lambda'}
~~, \label{total-s1loop-Left}
\eq
which is   HV, like the s-channel Born terms, but is numerically small
(100 to 1000 times smaller than Born, as expected from the $\alpha$ factor
occurring for an 1-loop correction without special enhancement effect). \\

{\bf (a3)  $H$ self-energy and right triangles.}\\
The implied amplitude  consists of two parts.
The  first part  $T_{\rm se}(SM)$  involves an initial $(\mu \mu H)$ coupling
followed by $H$ self-energy bubbles and a final $g_{HHH}$ coupling.
The second part, $T_{\rm tri}(SM)$, involves
an initial $(\mu \mu H)$ coupling followed by $H$ s-channel exchange and
either an SM $HHH$ form factor, or a bubble and a 4leg coupling to $HH$.
The sum of these 2 parts
\bq
T^{SM}_{HHH}(s)=T_{\rm se}(SM)+T_{\rm tri}(SM) ~~, \label{T1loopHHH}
\eq
leads to the helicity amplitude
\bq
F^{\rm HHH~SM}_{\lambda,\lambda'}=
{eg_{\mu\mu H}T^{SM}_{HHH}(s)\over s-m^2_H+im_H\Gamma_H}\sqrt{s}\delta_{\lambda,\lambda'}
~~, \label{total-s1loop-Right}
\eq
which, as (\ref{total-s1loop-Left}), is also HV and angle-independent.
The long list of contributions to $T^{SM}_{HHH}$ is collected in the Appendix.
It is found  that, above the $tt$ threshold, it is largely dominated by the $ttt$ contribution.
Numerically, the  amplitude in (\ref{total-s1loop-Right}) contains  important real and imaginary parts,
comparable  to the HV Born terms
in\footnote{Note that the SRS scheme \cite{SRS},  is globally satisfactory
above $\sim 1 {\rm TeV}$.}  (\ref{FBorn-helicity}).\\

We next turn to \underline{{\bf (b) in the $t,u$ sectors}} where we  find: \\

{\bf (b1) up or down triangles with $\mu^\mp$ exchange}:
\[
(\nu WW)~,~  (\mu ZZ)~,~ (\mu HH)~,~ (\nu GG)~,~ (\mu G^0G^0)~,~
(Z\mu \mu)~,~ (H\mu \mu)~,~ (G^0\mu \mu)~.~
\]
Among them, non negligible contributions only come from the $(\nu WW)$ and $(\mu ZZ)$ triangles,
which produce HV contributions (still 100 times weaker than
the Born ones) and  tending to an angular symmetric constant at high energy.
The other terms lead essentially to small HC amplitudes, like the corresponding Born ones
in (\ref{FBorn-helicity}).\\

{\bf (b2) $\mu^\mp$ self-energy bubbles}\\
These $ t$ and $u$ channel $\mu$ exchanges which were already small at Born
level lead to very small HC contributions when $\mu$ self-energy bubbles are added.\\

Finally, the  \underline{{\bf (c) Boxes}} are of two types: The direct boxes
\[
(\nu WWW)~,~ (\mu ZZZ)~,~ (\nu GGG)~,~ (\mu G^0G^0G^0)~,~ (\mu HHH)~,~
(Z\mu\mu\mu)~,~ (H\mu\mu\mu)~,~ (G^0\mu\mu\mu)~,~
\]
and the corresponding crossed ones; the twisted boxes
\[
(Z\mu\mu Z)~,~ (H\mu\mu H)~,~ (G^0\mu\mu G^0)~,~
\]
and the corresponding crossed ones.
Among them, the important contributions come from $(\nu WWW)$, $(\mu ZZZ)$
and their crossed boxes, because of the presence of $W,Z$ couplings
and the total absence of suppressed $H\mu\mu$ couplings. The induced invariant amplitudes
 are then given by
\bqa
 A_{\nu WWW}&=&-2\alpha^2g^2_{WWH}g^2_{W\mu L}\{-m_{\mu}D_{11}I_{1L}
+I_{1pL}(D_{12}-D_{13})\} ~,~ \nonumber \\
 A_{\mu ZZZ}&=&\alpha^2g^2_{ZZH}\Big \{4m_{\mu}g_{Z\mu L}g_{Z\mu R}I_1D_0
-2\Big [-m_{\mu}D_{11}(g^2_{Z\mu L}I_{1L}+g^2_{Z\mu R}I_{1R})\nonumber\\
&&+(g^2_{Z\mu L}I_{1pL}+g^2_{Z\mu R}I_{1pR}) (D_{12}-D_{13})\Big ]\Big \}~,~ \label{box-inv-amlpitudes}
\eqa
written in terms of Passarino-Veltman $D$-functions  \cite{PV},
and the helicity forms decomposed as
\bqa
I_{1}=\bar v(l',\lambda')P_L u(l,\lambda) & \to &
 -\sqrt{s}\delta_{\lambda,\lambda'} ~~,~  \nonumber\\
I_{1L}=\bar v(l',\lambda')P_L u(l,\lambda) &\to &
 -\sqrt{s}\delta_{\lambda,\lambda'}\delta_{\lambda,-} ~~,~ \nonumber\\
I_{1R}=\bar v(l',\lambda')P_R u(l,\lambda) &\to &
 -\sqrt{s}\delta_{\lambda,\lambda'}\delta_{\lambda,+} ~~,~ \label{HV-I-froms}
\eqa
for the  HV contributions which are suppressed by $m_{\mu}$ factors, and
\bqa
I_{1p}=\bar v(l',\lambda')\psl u(l,\lambda) & \to &
 -p_H\sqrt{s}\sin\theta\delta_{\lambda,-\lambda'}(\delta_{\lambda,-}
-\delta_{\lambda,+}) ~~,~~ \nonumber\\
I_{1pL}=\bar v(l',\lambda')\psl P_L u(l,\lambda) & \to &
 -p_H\sqrt{s}\sin\theta\delta_{\lambda,-\lambda'}\delta_{\lambda,-} ~~,~~ \nonumber\\
I_{1pR}=\bar v(l',\lambda')\psl P_R u(l,\lambda) & \to &
p_H\sqrt{s}\sin\theta\delta_{\lambda,-\lambda'}\delta_{\lambda,+}~~,~ \label{HC-I-froms}
\eqa
for  the  HC contributions. In (\ref{box-inv-amlpitudes}), the couplings
\bqa
&& g_{W\mu L}={1\over s_W\sqrt2}~~,~~ g_{Z\mu L}=-{1-2s^2_W\over 2s_Wc_W}~~,~~
g_{Z\mu R}={s_W\over c_W} ~~, \nonumber \\
&& g_{WWH}={m_W\over s_W}~~, ~~ g_{ZZH}={m_Z\over s_Wc_W}~~,~~ \label{couplings2}
\eqa
are used. Notice that the HC amplitudes of these boxes do not involve any mass
suppressed coupling constant, such that they are only reduced by the one loop
$\alpha/\pi$ factor. For comparison the  Born HC amplitudes
(due to $t,u$ channels $\mu^\mp$ exchange) are reduced by the factor $(m_\mu/ m_W)^2$.
Due to these features the box HC amplitudes are
$({\alpha\over\pi})({m_W\over m_{\mu}})^2\sim 1000$ times larger then
the HC Born ones and have a size almost comparable to the Born HV amplitudes. Both real
and imaginary parts are important.

The remaining box HV contributions coming from $m_{\mu}$ terms do not
have this enhancement and get the usual $\alpha/\pi$ reduction factor
as compared to the Born terms.\\

{\bf Resulting total SM amplitudes and cross sections.}\\
These are constructed by summing  the various 1-loop contributions
and adding them to the Born ones. The important terms are the HV Born term  of (\ref{FBorn-helicity}),
the H self-energy and HHH right triangles of (\ref{total-s1loop-Right})
 coming essentially from top quark diagrams
(see Appendix), and the aforementioned  contribution from the two boxes leading to
 (\ref{box-inv-amlpitudes}).

These final complete 1-loop results (Born + 1-loop diagrams)
for the HC and HV SM helicity amplitudes are illustrated in Fig.1,
where we show their energy and angle dependencies.
Note that these complete SM HC and HV amplitudes have  comparable sizes,
although their $s$- and $\theta$-dependencies are rather different.
In particular
the two HC amplitudes almost coincide at 0.5 TeV (upper right panel),
which also happens for the HV amplitudes at $\theta=60^\circ$,
in a wide range of energies (lower left panel).\\

The corresponding  (complete 1-loop) cross sections, shown in Fig.2, reflect the  properties
of the unpolarized cross sections defined in (\ref{dsigma-unpol}),
 and the $\mu^\mp$ polarized cross sections denoted as $d\sigma_{\lambda \lambda'}/d\cos\theta$.
For the later, the alternative notation indicating  whether the contributing amplitudes
are HV or HC   may also be inspiring:
\bqa
&& d\sigma_{LL} \equiv d\sigma_{L}(HV) ~,~ d\sigma_{RR} \equiv d\sigma_{R}(HV) ~,~
d\sigma(HV)\equiv d\sigma_L(HV)+d\sigma_R(HV) ~,~\label{HV-sigma} \\
&& d\sigma_{LR}\equiv d\sigma_{L}(HC) ~,~ d\sigma_{RL}\equiv d\sigma_{R}(HC) ~,~
d\sigma(HC)\equiv d\sigma_L(HC)+d\sigma_R(HC) ~.~\label{HC-sigma}
\eqa
As seen in Fig.2 the HV differential cross sections are  angularly constant at $\sqrt{s}=0.5$TeV
and reflect mainly the s-channel $H$ exchange parts containing the effect of the HHH form factor.
On the contrary, the HC cross section
has a specific angular shape; it starts with smaller
values at low energy, but becomes of comparable size to the HV one at high energies,
due to the large box contributions.

The unpolarized cross section contains all these features but
obviously does not allow their easy disentangling.\\

\section{Anomalous HHH contributions}

We are now looking for possible effects of a modification
of the SM $HHH$ form factor due to new physics contributions.
They will only affect the HV helicity amplitudes according to
the $H$ exchange diagram, giving
\bq
\delta F_{\lambda,\lambda'}=
{eg_{\mu\mu H}\delta T_{HHH}(s)\over s-m^2_H+im_H\Gamma_H}\sqrt{s}\delta_{\lambda,\lambda'}~~,
\eq
where $\delta T_{HHH}(s)$ is the departure to the SM prediction $T^{SM}_{HHH}(s)$ in (\ref{T1loopHHH}),
which is induced by the $H$ self-energy and $HHH$ form factor discussed in the {\bf a3} part
of Section 3. \\

{\bf Examples of new physics contributions}\\
Our aim is not to study particular new models but only to look
at the sensitivity of the $\mu^-\mu^+\to HH$ process to modifications
of the SM $g_{HHH}$ coupling and especially to the $s$ dependent
form factor that they generate.

Modifications are often described by effective operators,
see \cite{Anom1,Anom2}. There are various types of dimension-6 operators
leading to anomalous couplings. Among them we can mention the ones generating direct effective
$\mu\mu HH$ couplings like
\bq
O={c\over\Lambda^2}(H^\dagger  \overleftrightarrow{D_{\mu}}H)(\bar \mu\gamma^{\mu}P_{L,R}\mu)
\eq
\noindent
which would add global contributions to the
$\mu^-\mu^+\to HH$ amplitude of the type
\bq
c{s\over\Lambda^2} v(l',\lambda')(\ppsl-\psl)P_{L,R} u(l,\lambda)
\eq
as long as $s\ll\Lambda^2$, where $\Lambda$ is an effective new physics scale .\\
More closely related to our study of the $HHH$ form factor there is also a dimension 6 operator
\bq
O_T={c_T\over2\Lambda^2}(H^\dagger  \overleftrightarrow{D_{\mu}}H)^2 ~~,
\eq
\noindent
which would give an additional contribution $\delta T_{HHH}(s)$ to the standard HHH coupling
of the type $c_T(s/\Lambda^2)$.\\

However these descriptions only parameterize a departure from the SM prediction
as long as $s\ll\Lambda^2$ but not a  complete $s$ dependence (the shape)
of the form factor which is our purpose.

We therefore come back to the precise structure of the $HHH$ vertex.
With the idea of compositeness we can take as example the hadronic
structure of the $\sigma\sigma\sigma$ vertex where the $\sigma$
is a $q\bar q$ bound state. This vertex can be pictured through
a triangular quark loop, but it is obviously affected by non perturbative
binding interactions.
With such a picture the whole $HHH$ coupling should then come
from $(XXX)$ triangles made by the constituents $X$ and an effective
$HXX$ coupling related to the binding.
This would generate an effective $HHH$ vertex replacing
the usual SM $HHH$ Born term.
On another hand, if the Higgs boson is connected to a new sector, one may have
triangles involving the corresponding new particles.
In the case of a strong sector (similarly to the hadronic case),
there may be resonances $R$ leading to $HHH$ contributions
of the type $H\to R(XX) \to HH$.

In Fig.3 we give illustrations of the contributions to the $(HHH)$ form factor
corresponding to such examples and we compare them to the total SM one
(essentially controlled by the $ttt$ triangle) and to its supersimplicity approximation
(called sim) given in the {\bf a3} part of Section 3.

\begin{itemize}

\item
For a scalar Higgs-constituent $X$ with a $g_{HXX}$ coupling,
we get the departure $\delta T_{HHH}(s)$
due to the $XXX$ triangle contribution to the $HHH$ coupling:
\bqa
&&\delta T_{HHH}(s)\to A_{XXX}(s)=-~{e\alpha\over4\pi}g^3_{HXX}
C_0(s,m^2_H,m^2_H,m^2_X,m^2_X,m^2_X)
\eqa
where $C_0$ is the Passarino-Veltman \cite{PV} function.
Using its high energy expansion \cite{asPV} one gets
(see (A.1))
\bqa
 -~{e\alpha\over4\pi}g^3_{HXX}{ \overline{ \ln^2s_X}\over 2s} ~~, \label{ANPXXX}
\eqa
In the illustration we take $m_X=0.5$ TeV and $g_{HXX}=-10$ TeV.

\item
for a fermionic constituent $F$, we get similarly the departure due to the $FFF$ triangle
\bqa
&&\delta T_{HHH}(s)\to A_{FFF}(s)=-~{e\alpha\over4\pi}g_{HFF}^3
\Big \{2m^3_FC_0+2m_F \Big [3m^2_H(C_{21}+C_{22})+6p.p'C_{23}\nonumber\\
&& +3nC_{24}
+2q.pC_{11}+2q.p'C_{12}+2m^2_HC_{11}+2p.p'C_{12}+q.pC_{0}\Big ]\Big \}
\nonumber\\
&&\to -~{e\alpha\over4\pi}g_{HFF}^3
\{2m_F[{- \overline{ \ln^2s_F} \over4}- \overline{ \ln s_{FF}}]\}~~, \label{ANPFFF}
\eqa
with the notations defined in eq.(A.1,A.2) of the Appendix.
In the illustration we take $m_F=0.5$ TeV and $g_{HFF}=-4$.

\item

and for a typical resonance effect  we get the (trivial) shape
\bq
\delta T_{HHH}(s)\to A_{Res}(s)= {g_{HR}g_{RHH}\over s-M^2_R+iM_R\Gamma_R} ~~, \label{ANPRes}
\eq
The illustration is made with   $M_R=1$ TeV, $\Gamma_R=0.3$ TeV, $g_{HR}g_{RHH}=0.5$ TeV. \\
\end{itemize}

The numerical values of the above masses and effective couplings have been  chosen such that,
in the illustrations, the shape of the resulting $HHH$ form factor can be quickly
compared with that of the SM case such that one can
appreciate the different spectacular $s$-dependencies. One indeed sees that the $s$ dependencies
appearing in these examples are very different from each other and also very different from the SM
case.

So we believe that there is much to learn  from the measurement of the $HHH$ form
factor.\\

 We can now see how this reflects in the HV
$\mu^-\mu^+\to HH$ amplitudes (see Fig.4)
and in the cross sections (see Fig.5)
with their specific energy dependencies, threshold effects and resonance shapes.

In Fig.5 we show the relative differences $[\sigma_{SM+NP}-\sigma_{SM}]/\sigma_{SM}$
between the cross sections involving these new contributions and the pure SM cross sections,
for the HV contribution (see (\ref{HV-sigma})) and for unpolarized case.
Due to the common dominating final $\delta T_{HHH}(s)$ term,
the Left-Left and Right-Right HV cases defined in (\ref{HV-sigma}) would give
similar results to the complete HV case.
So one sees that polarized beams allowing the separation of HV from HC contributions
would help for differentiating $HHH$ form factor effects from possible other anomalous effects.

As one can see in the illustrations, the $s$ and $\theta$ dependencies
of the cross sections (even the unpolarized one) should allow to identify
the nature of the new contribution.

At this point we should  add a few words about the observability of such effects.
The energy of a $\mu^-\mu^+$ collider has been considered up to
6 TeV \cite{Delahaye}. For our study the required energy would correspond to the
yet unknown new physics scale, although the curious anomalies
observed at LHC around 0.75 and 2 TeV \cite{bumps} could be in mind
but it is too early to know how they would affect the $HHH$ coupling.\\
With an expected luminosity of the order of $10^{35} cm^{-2}s^{-1}$
\cite{Delahaye}, the SM cross section (see Fig.2) would lead to only
a few events per year.
But we have seen that large enhancements could appear due to anomalous
$HHH$ couplings, threshold and resonance effects in the $HHH$ form factor,
which should then be observable.
In case these luminosities could not be reached,
we can mention that there may be other processes
(for example WW fusion, see next Section)
where $HH$ production could be observed with a higher statistics, see for ex
\cite{Conway}.\\

\section{Conclusions}

In this work we have  computed the full 1-loop SM contributions to the
$\mu^-\mu^+\to HH$ process and we have studied the role
of the final $HHH$ coupling and of its SM form factor.
Our aim is to show how possible new contributions to
the $HHH$ form factor could be identified through  observables.
We have emphasized the specific properties of the HC, HV amplitudes,
their energy and angular dependencies and how this reflects in
the observable polarized and unpolarized cross sections.

We have compared the real and imaginary parts of the  SM 1-loop contributions
to the $HHH$ form factor, to examples of possible new physics effects
corresponding either to
Higgs boson compositeness or to the coupling of the Higgs boson to a new sector.
In each case we have also given the corresponding simple analytic expressions,
in the adequate "sim" approximation discussed in the Appendix and
 allowing a quick estimate of the effect at high $s$.

We have emphasized the fact that the $q^2\equiv s$ dependencies
of the $HHH$ form factor can be very different, depending on
their origin.  We have taken some arbitrary cases with new scalar or new fermion
contributions to the $HHH$ form factor, or strong resonances,
and made the corresponding illustrations.
As it can be seen in these illustrations,
the differences can be spectacular and reflect the specific nature
of the new physics.

We have shown that polarized cross sections ($\mu^{\mp}$
beam polarization could be available according to some studies
 \cite{mumucol, mumupol}) are essential
for differentiating HV contributions (which are the only ones
containing $HHH$ form factor effects) from HC contributions.

But even the shape of the $s$- and $\theta$-dependencies of the unpolarized
cross sections should help for identifying the nature of the new contribution.

The present study is an example of what could be done
for the search of $HHH$ form factor effects in the process $\mu^-\mu^+ \to HH$.
Spectacular resonance or threshold effects could easily be seen but high
luminosity would be required in order to make precise analyses.
This would correspond to the simplest situation.

More complex processes like $ZZ \to HH$, $W^-W^+ \to HH$,
$gg\to HH$ or $\gamma\gamma \to HH$ could be considered
and would benefit of larger cross sections at $e^-e^+$, $\mu^-\mu^+$  colliders or at LHC.
Note that the fusion subprocesses $ZZ \to HH$ and $W^-W^+ \to HH$
involve, like in the above $\mu^-\mu^+$ case, the simple s-channel $H$ exchange diagram,
with in addition a 4-leg $ZZHH$ or $W^-W^+HH$ vertex, as well as $t$ and $u$ channel
$Z$ or $W$ exchanges. These subprocesses can be measured by making detailed specific analyses.

The processes  $gg \to HH$ and $\gamma\gamma \to HH$
contain an s-channel $H$ exchange, but the initial vertex needs
a 1-loop contribution and there are also several other 1-loop
diagrams producing the final $HH$ state.
Specific works should be devoted to each of these processes,
see e.g.\cite{ggHH, Dawson, Telnov, Asakawa, Levy}.

The aim of this paper was only to to put forward the idea of looking
especially at the $s$-dependence of the $HHH$ form factor and to show
that observable consequences may exist.

We hope that these first results will encourage further phenomenological
and experimental studies of the possibilities to measure this form
factor.\\


\vspace*{2.cm}

\renewcommand{\thesection}{}

\section{Appendix: SM contributions to the   $HHH$\\ form factor.}

\setcounter{equation}{0}
\setcounter{subsection}{0}
\renewcommand{\thesubsection}{A.\arabic{subsection}}
\renewcommand{\theequation}{A.\arabic{equation}}

The SM prediction for the $HHH$ form factor  consists of a zero order contribution
given by the  point-like coupling $g_{HHH}$ in (\ref{couplings1}), and of higher
order corrections.
In the OS scheme \cite{OS, HHHOS} these corrections consist of parameter renormalization
and additional 1-loop diagrams.
We are interested in the $q^2\equiv s$ dependence when one $H$ is off-shell,
while the two other $H$ (with four-momenta $p,p'$) are on-shell, in order to make
the comparison of this SM prediction with possible compositeness structures.

So we will use a procedure allowing to quickly get simple forms which reflect sufficiently
well the size and the $s$-dependence of each contribution.
This is the supersimple renormalization scheme
(SRS) procedure \cite{SRS, WW}, which
leads to the simplest expressions in terms of augmented Sudakov logarithms.
Among them,  we will only need the augmented Sudakov forms (see \cite{SRS,WW} for details):
\bqa
&& \overline{\ln^2s_X}\equiv
\ln^2 s_X +4L_{HXX} ~~, ~~~
s_X \equiv \left (\frac{-s-i\epsilon}{m_X^2} \right ) ~~, \label{Sud-ln2} \\
&& \overline{\ln s_{ij}}\equiv \ln s_{ij}+b^{ij}_0(m_H^2)-2 ~~,~~~
\ln s_{ij}\equiv \ln \frac{-s-i\epsilon}{m_im_j} ~~, \label{Sud-ln}
\eqa
 where $(X,i,j)$ refer to internal exchanges in the contributing diagrams. The
   explicit expressions for $b_0^{ij}(m^2_H)$ and $L_{HXX}$
 are given e.g. in  (A.6, A.5) of \cite{WW}.
 We note that the counter terms needed
 in the SRS scheme  respect the  structure (\ref{Sud-ln2}, \ref{Sud-ln}) \cite{SRS,WW}.

Globally this procedure consists in replacing the divergent terms
related to the  $(i,j)$ internal lines of any contributing   diagram,   as
\bq
\ln \frac{-s-i\epsilon}{\mu^2} -\Delta \to \ln s_{ij}+b_0^{ij}(m^2_H)
~~, \label{SRS-replacement}
\eq
where $\mu$ here denotes the renormalization scale and $\Delta=1/\epsilon -\gamma_E +\ln (4\pi)$,
with  the number of dimensions used for regularization written as $n=4-2\epsilon$.

In the present case, with only  triangle and bubble diagrams contributing,
there is no ambiguity related to the internal lines $(i,j)$.
They can only be  $H$, $Z$, $W$ and $t$, so that we can only have $(ij)=(HH),(ZZ),(WW), (tt)$.
The SRS results thus obtained, are always denoted as ``sim"
in the illustrations \cite{SRS,WW}.\\

We next describe  the exact   expressions for the various
triangle and bubble diagrams with 4-leg couplings,  as well as
 their high energy SRS (sim) forms. At first $\alpha$ order,
 the $T^{SM}_{HHH}(s)$  form factor of (\ref{T1loopHHH}), may be  written as
\bq
T^{SM}_{HHH}(s)=eg_{HHH}+A^{SM}(s) ~~. \label{HHH-ASM}
\eq
In the following two subsections we first give the  $A^{SM}(s)$ results implied by
the 1-loop triangles and bubbles with a 4-leg couplings, and then from the  $H$ self-energy.

\subsection{Triangles and bubbles with 4-leg couplings}

Depending on the natures of the exchanged particles,
the contributions to $A^{SM}(s)$ from the various  2-point and 3-point
Passarino-Veltman functions, denoted as B and C,   are given by  \cite{PV, asPV}: \\

\begin{itemize}
\item

\noindent
{\bf Scalar $(SSS)$ triangles  and $(SS)$ bubbles with a 4-leg $SSHH$ coupling.}
\bqa
A^{SM}_{SSS}(s) &=& -{e\alpha\over4\pi}\Bigg \{g^3_{HSS}
C_0(s,m^2_H,m^2_H,m^2_S,m^2_S,m^2_S) \nonumber \\
 && +g_{HSS}g_{HHSS}\Big [B_0(s,m^2_S,m^2_S)+2B_0(m^2_H,m^2_S,m^2_S \Big ] \Bigg \}\nonumber\\
& \to & -~{e\alpha\over4\pi}\Bigg  \{g^3_{HSS}{ \overline{ \ln^2s_S}\over s}
- g_{HSS}g_{HHSS} ~ \overline{\ln s_{SS}} \Bigg \}~~. \label{ASMSSS}
\eqa
This applies to the triangles
\[
SSS = HHH, G^0G^0G^0, C^ZC^ZC^Z, G^{\pm}G^{\pm}G^{\pm}, C^{\pm}C^{\pm}C^{\pm}  ~,
\]
and  the bubbles
\[
 SS = HH, G^0G^0, G^{\pm}G^{\pm}  ~,
\]
while  $g_{HHH}$  is given in (\ref{couplings1})
and\footnote{$(G^\pm, G^0)$ denote the SM Goldstone fields and $(C^\pm, C^Z)$ the
FP ghosts.}
\bqa
&& g_{HHHH}=-~{3m^2_H\over 4s^2_Wm^2_W} ~,~
g_{HGG}=-~{m^2_H\over 2s_Wm_W}  ~,~  g_{HHGG}={1\over 2s^2_Wc^2_W} ~, \nonumber \\
&& g_{HC^ZC^Z}=-~{m_W\over 2s_Wc^2_W} ~,~ g_{HC^{\pm}C^{\pm}}=-~{m_W\over 2s_W} ~. \label{couplings3}
\eqa
Note that there is no 4-leg diagram for the ghost loop,  and that
 a global fermionic minus sign has been inserted.
In all cases, the internal $S$ mass
for $H, G^0, C^Z, G^\pm, C^\pm$ is respectively equal to the one of $H,Z,Z,W, W$.\\

\item

\noindent
{\bf Fermion triangles $(fff)$.}\\
Due to the strong mass dependence of the $Hff$ coupling,
it is adequate to restrict to the $(ttt)+(\bar t\bar t\bar t)$ case. The result is
\bqa
&& A^{SM}_{ttt}(s) = {e\alpha\over4\pi} {3m^3_t\over2s^3_Wm^3_W}
\Bigg \{2m^3_tC_0+2m_t[3m^2_H(C_{21}+C_{22})+6p.p'C_{23}\nonumber\\
&& +3nC_{24} +2q.pC_{11}+2q.p'C_{12}+2m^2_HC_{11}+2p.p'C_{12}+q.pC_{0}]\Bigg \}
\nonumber\\
&& \to ~{e\alpha\over4\pi} {3m^3_t\over2s^3_Wm^3_W}
\Bigg \{2m_t \Bigg [{- \overline{\ln^2s_t} \over4}- \overline{ \ln s_{tt}} \Bigg ] \Bigg \}
~~. \label{ASMttt}
\eqa\\

\item
\noindent
{\bf Vector triangles $(VVV)$ and bubbles $(VV)$  with a 4-leg $HHVV$ coupling.}
\bqa
&&A^{SM}_{VVV}(s)={e\alpha\over4\pi} \Big \{g^3_{HVV}nC_0
+g_{HVV}g_{HHVV}[2B_0(m^2_H,m^2_V,m^2_V)+B_0(s,m^2_V,m^2_V)] \Big \}
\nonumber\\
&&\to -~{e\alpha\over4\pi} \Big \{2g^3_{HVV}{\overline{\ln^2s_V}\over s}
+g_{HVV}g_{HHVV}[- \overline{\ln s_{VV}}] \Big \} ~~, \label{ASMVVV}
\eqa
applied only to $V=Z,~W$, since there are no $HZ\gamma$ or $HHZ\gamma$ couplings.
Because of this, the $V$ masses in the SRS forms  $\overline{ \ln^2s_V}$ and $\overline{ \ln s_{VV}}$
 can either be   $m_Z$ or  $m_W$.\\

\item
\noindent
{\bf $(VVS)$ triangles}
\bqa
&&A^{SM}_{VVS}(s)={e\alpha\over4\pi}g^2_{VSH}g^2_{VVH} \Big \{
m^2_H(C_{21}+C_{22})+2p.p'C_{23}+nC_{24}
\nonumber\\
&& +(p.p'+3q.p)C_{11}+(m^2_H+3q.p')C_{12}+2(q^2+q.p')C_{0}\Big \}
\nonumber\\
&&\to {e\alpha\over4\pi}g^2_{VSH}g^2_{VVH}\Big \{
{1\over2}( \overline{\ln^2s_V}+ \overline{\ln s_{VV}}) \Big \}~~, \label{ASMVVS}
\eqa
applied to $ZZG^0, ~ W^{\pm}W^{\pm}G^{\pm}$; compare (\ref{ASMVVV}).\\

\item

\noindent
{\bf $(VSV)$ triangles}
\bqa
&&A^{SM}_{VSV}(s)={e\alpha\over4\pi}g^2_{VSH}g^2_{VVH}\Big \{
m^2_H(C_{21}+C_{22})+2p.p'C_{23}+nC_{24}
\nonumber\\
&& +(3m^2_H-p.p')(C_{11}-C_{12})+2(m^2_H-p.p')C_{0} \Big \}
\nonumber\\
&&\to {e\alpha\over4\pi}g^2_{VSH}g^2_{VVH} \Big \{
{1\over4} \overline{ \ln^2s_V}+2 \overline{ \ln s_{VV}} \Big \}~~, \label{ASMVSV}
\eqa
applied to $ZG^0Z, ~W^{\pm}G^{\pm}W^{\pm}$.\\

\item
\noindent
{\bf $(SVV)$ triangles}
\bqa
&&A^{SM}_{SVV}(s)={e\alpha\over4\pi}g^2_{VSH}g^2_{VVH}\Big \{
m^2_H(C_{21}+C_{22})+2p.p'C_{23}+nC_{24}
\nonumber\\
&& -(m^2_H+q.p)C_{11}-(p.p'+q.p')C_{12})+q.pC_{0} \Big \}\nonumber\\
&&\to {e\alpha\over4\pi}g^2_{VSH}g^2_{VVH}\Big \{
-{1\over2}( \overline{ \ln^2s_V}+ \overline{ \ln s_{VV}})\Big \}~~, \label{ASMSVV}
\eqa
applied to $G^0ZZ, G^{\pm} W^{\pm}W^{\pm}$.\\

\item
\noindent
{\bf $(VSS)$ triangles}
\bqa
&&A^{SM}_{VSS}(s)={e\alpha\over4\pi}g^2_{VSH}g^2_{SSH}\Big \{
-m^2_H(C_{21}+C_{22})-2p.p'C_{23}-nC_{24}
\nonumber\\
&& -2(m^2_H+q.p)C_{11}-2(p.p'+q.p')C_{12})-4q.pC_{0}\Big \}
\nonumber\\
&&\to {e\alpha\over4\pi}g^2_{VSH}g^2_{SSH}\Big \{
-{1\over2} \overline{ \ln^2s_V}+ \overline{ \ln s_{VV}}\Big \}~~, \label{ASMVSS}
\eqa
applied to $ZG^0G^0, W^{\pm}G^{\pm} G^{\pm}$.\\

\item
\noindent
{\bf $(SVS)$ triangles}
\bqa
&&A^{SM}_{SVS}(s)={e\alpha\over4\pi}g^2_{VSH}g^2_{SSH}\Big \{
-m^2_H(C_{21}+C_{22})-2p.p'C_{23}-nC_{24}
\nonumber\\
&& -(-m^2_H+q.p+p.p')C_{11}-(m^2_H-p.p'+q.p')C_{12})+(p.p'+q.p)C_{0}\Big \}
\nonumber\\
&&\to {e\alpha\over4\pi}g^2_{VSH}g^2_{SSH}
\{ \overline{ \ln^2s_V}- \overline{ \ln s_{VV}}\}~~, \label{ASMSVS}
\eqa
applied to $G^0ZG^0, ~G^{\pm}W^{\pm} G^{\pm}$.\\

\item
\noindent
{\bf $(SSV)$ triangles}
\bqa
&&A^{SM}_{SSV}(s)={e\alpha\over4\pi}g^2_{VSH}g^2_{SSH} \Big \{
-m^2_H(C_{21}+C_{22})-2p.p'C_{23}-nC_{24}
\nonumber\\
&& -(m^2_H-q.p-p.p')C_{11}-(-m^2_H+p.p'-q.p')C_{12})+(-q.p'+q.p)C_{0}\Big \}
\nonumber\\
&&\to {e\alpha\over4\pi}g^2_{VSH}g^2_{SSH}\Big  \{
-{1\over2} \overline{ \ln^2s_V}+ \overline{ \ln s_{VV}}\Big  \}~~, \label{ASMSSV}
\eqa
applied to $G^0G^0Z, ~ G^{\pm}G^{\pm}W^{\pm}$.\\

\end{itemize}

In the above contributions the following couplings are needed
\bqa
&& g_{ZZH}={m_Z\over s_Wc_W} ~,~   g_{ZZHH}={1\over 2s^2_Wc^2_W} ~,~ \nonumber \\
&& g_{WWH}={m_W\over s_W}  ~,~  g_{WWHH}={1\over 2s^2_W} ~,~
 g_{ZGH}=g_{WGH}={1\over 2s_Wc_W} ~~ . \label{couplings4}
\eqa\\

\subsection{ $H$ self-energy}

This additional contribution is given by
\bq
A^{SM}_{se}(s)=-~{eg_{HHH}\over s-m^2_H}\Sigma_H(s)~~, \label{ASMse}
\eq
where $\Sigma_H(s)$ is computed from the following diagrams:\\

\begin{itemize}

\item
\noindent
{\bf Bubbles $VV$ } leading to
\bq
\Sigma_H(s)
={X^2_1\over4\pi^2}[B_0]\to
{X^2_1\over4\pi^2}[- \overline{ \ln s_{VV}}] ~~,\label{SigmaHVV}
\eq
for which we respectively get
\bqa
VV=ZZ &\to &  X^2_1={e^2M^2_W\over 2s^2_Wc^4_W} ~~, \nonumber \\
VV=W^{\pm}W^{\mp} &\to &  X^2_1={e^2M^2_W\over s^2_W}~~. \label{VVX2}
\eqa\\

\item
\noindent
{\bf Bubbles $SV$ } leading to
\bqa
\Sigma_H(s)
 && =-~{X^2_1\over16\pi^2}[s(B_0+B_{21}-2B_1)+nB_{22}]
 \nonumber\\
&& \to -~{X^2_1\over16\pi^2}[-2s \overline{\ln s_{SV}}]~~,\label{SigmaHSV}
\eqa
for which we respectively get
\bqa
SV=G^0Z  &\to & X^2_1={e^2\over4s^2_Wc^2_W}~~, \nonumber \\
SV=G^{\mp}W^{\pm} &\to &  X^2_1={e^2\over2s^2_W}~~. \label{SVX2}
\eqa\\

\item
\noindent
{\bf \underline{Bubble $tt$}} leading to
\bqa
&&\Sigma_H(s)
=-~{1\over4\pi^2}[(s(B_1+B_{21})+nB_{22}+m^2_tB_0)X^2_1]\nonumber\\
&&\to
-~{X^2_1\over4\pi^2}[{s\over2} \overline{\ln s_{tt}}]~~,\label{SigmaHtt}
\eqa
with
\bq
X^2_1={3e^2\over 4s^2_WM^2_W}[m^2_t]~~. \label{ttX2}
\eq\\

\item
\noindent
{\bf Bubbles $SS$} leading to
\bq
\Sigma_H(s)
={X^2_1\over16\pi^2}[B_0] \to
{X^2_1\over16\pi^2}[- \overline{\ln s_{SS}}]~~,\label{SigmaHSS}
\eq
with
\bq
X^2_1 = {9e^2m^4_H\over8s^2_WM^2_W}~,~
{e^2m^4_H\over8s^2_WM^2_W}~,~{e^2m^4_H\over4s^2_WM^2_W}~,~
-{e^2m^2_W\over4s^2_Wc^4_W}~,~-{e^2m^2_W\over2s^2_W}~, \label{SSX2}
\eq
for
\bq
SS=~~HH,~~G^0G^0 ~,~ G^+,G^- ~,~ C^ZC^Z~,~ C^+C^-~~, \label{SSX2a}
\eq
respectively. Note that in these $SS$ bubbles, the internal $S$ mass is correspondingly
equal to the mass of $H,Z,W,Z,W$.\\

\end{itemize}

\begin{figure}[b]
\[
\epsfig{file=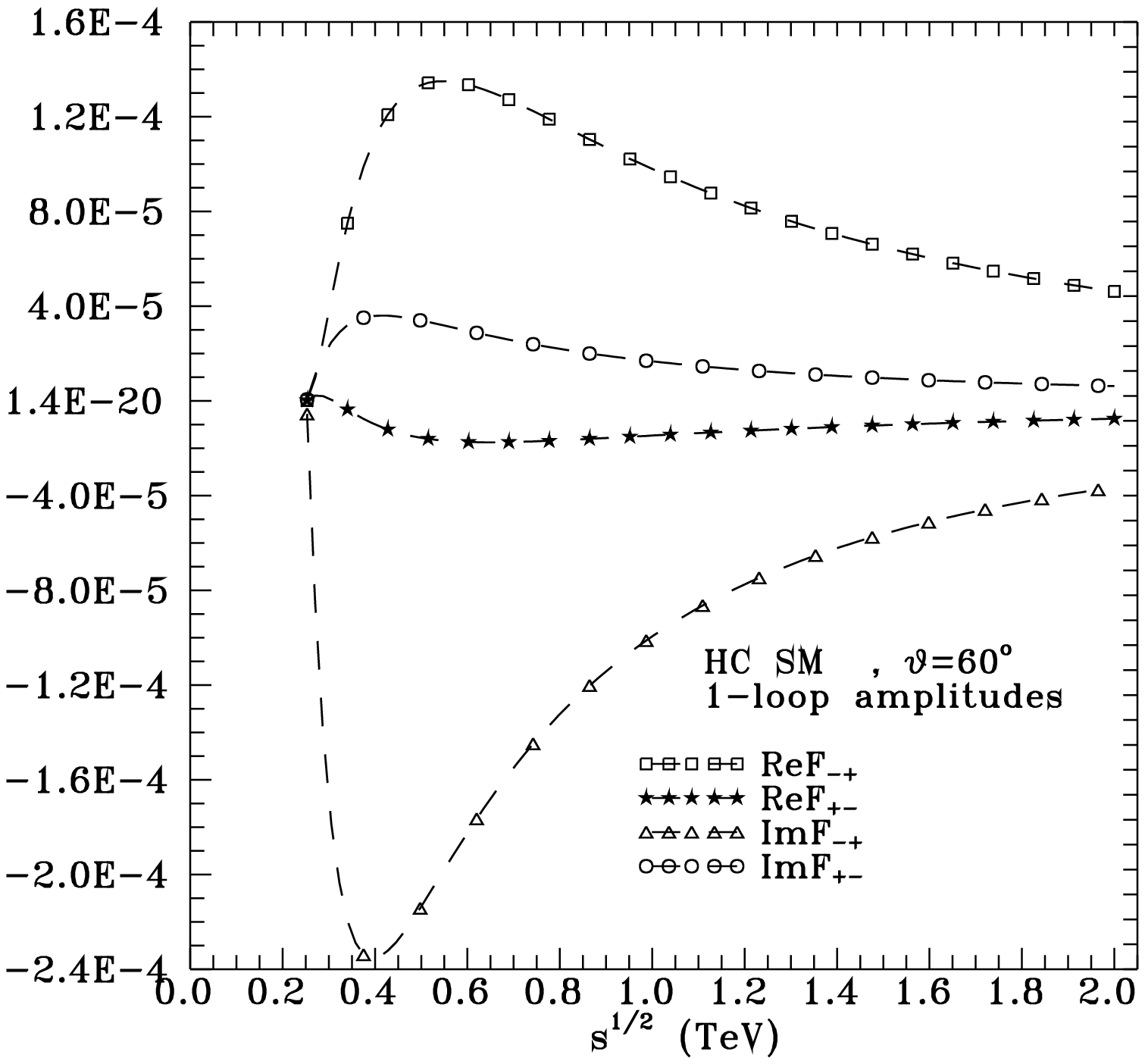, height=7.cm}\hspace{0.5cm}
\epsfig{file=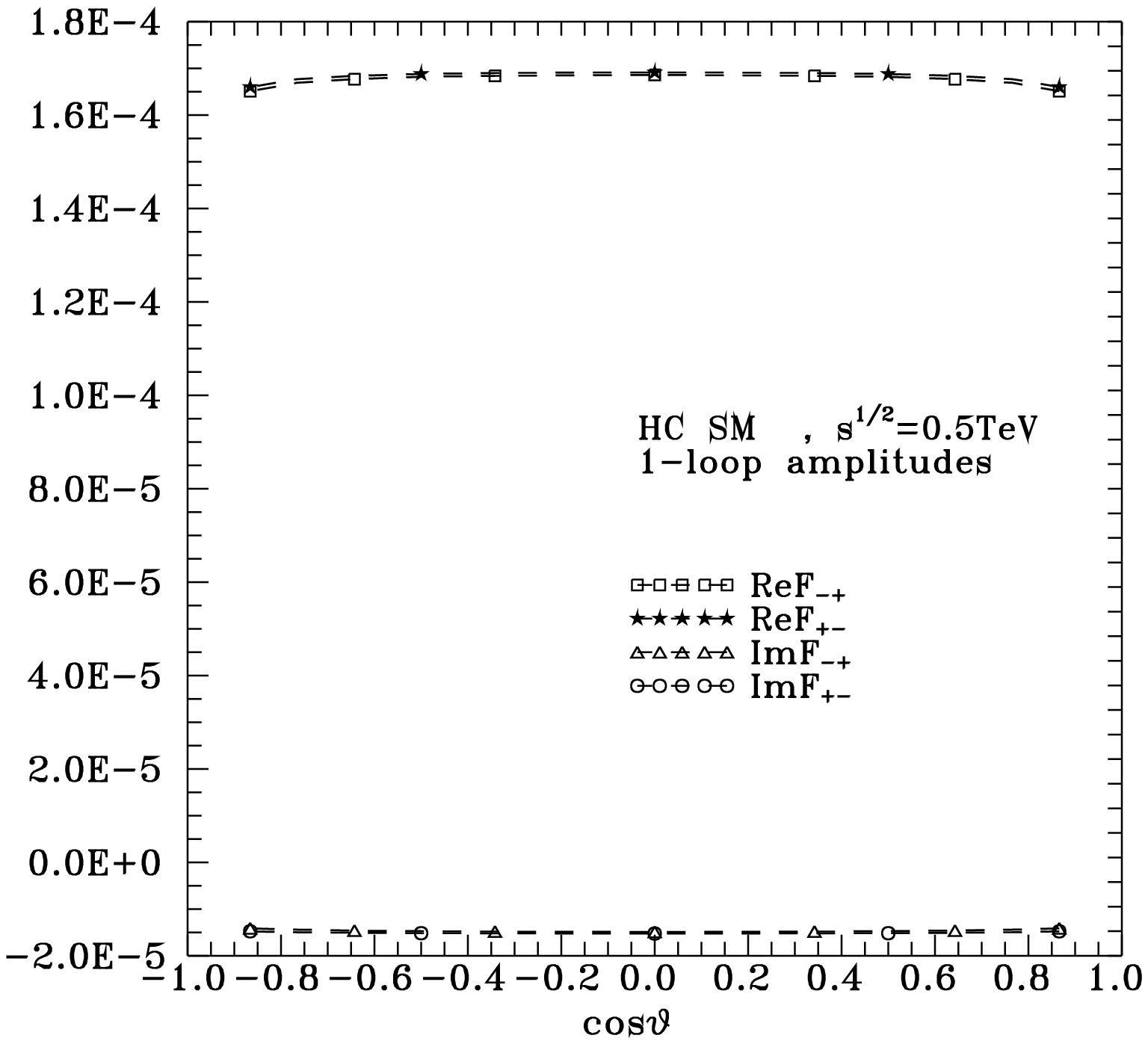,height=7.cm}
\]
\[
\epsfig{file=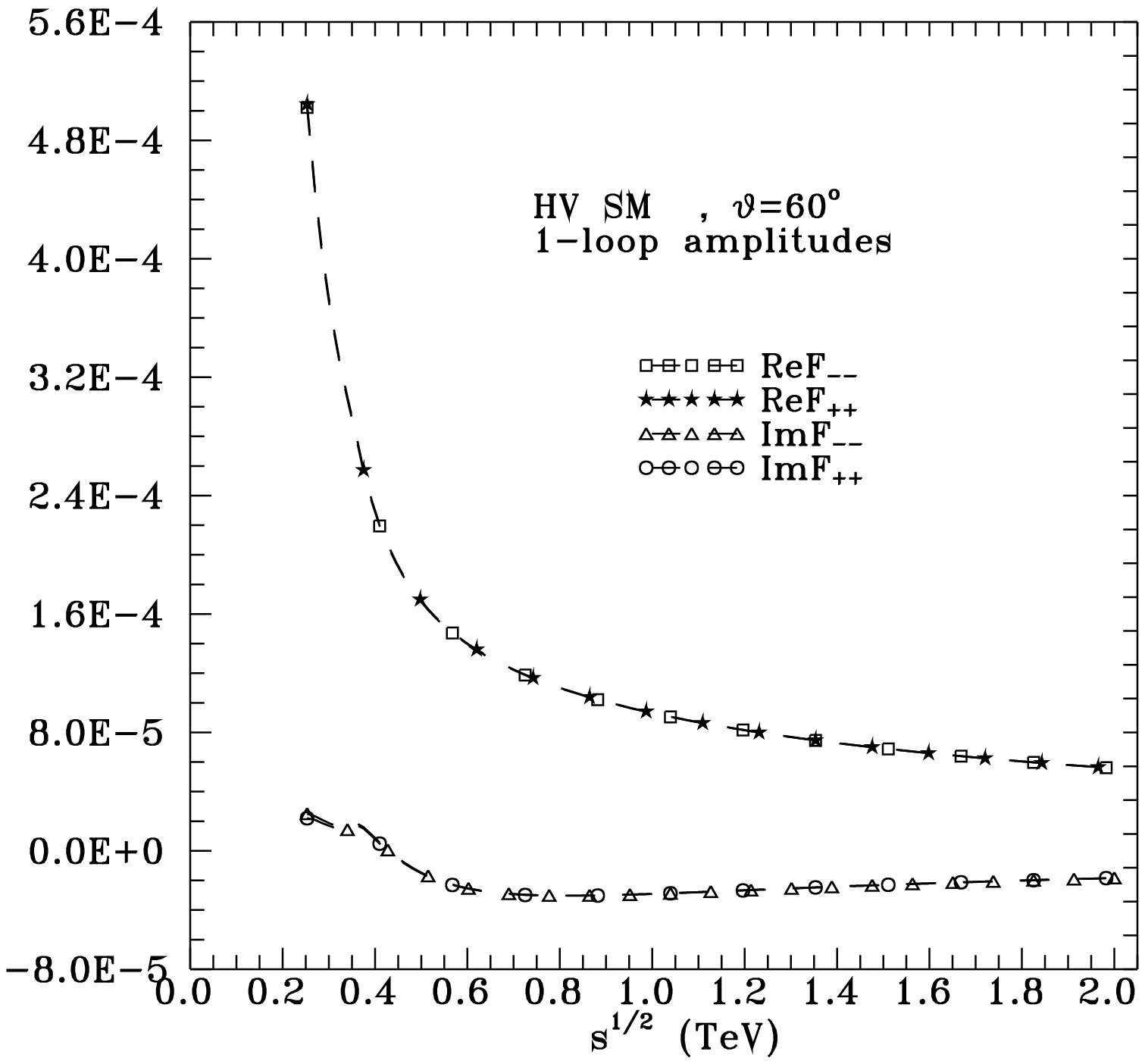, height=7.cm}\hspace{0.5cm}
\epsfig{file=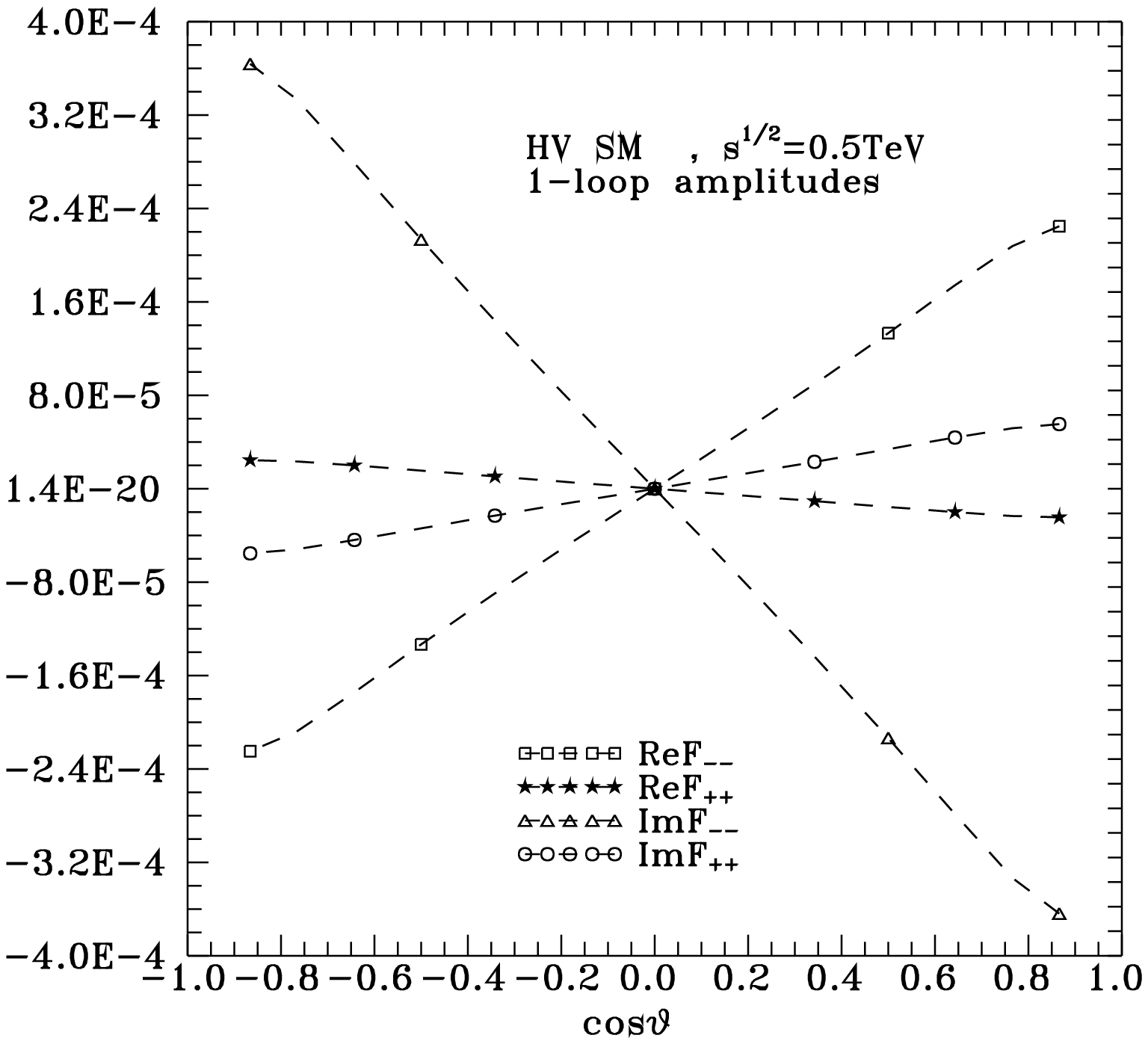,height=7.cm}
\]
\caption[1]{The complete SM 1-loop HC amplitudes (upper panels) and the corresponding
 HV amplitudes (lower panels).
 Left panels present the energy dependencies at $\theta=60^\circ$, while right panels
the angular dependencies at $\sqrt{s}=0.5 {\rm TeV}$. }
\label{F1g1}
\end{figure}

\begin{figure}[b]
\[
\epsfig{file=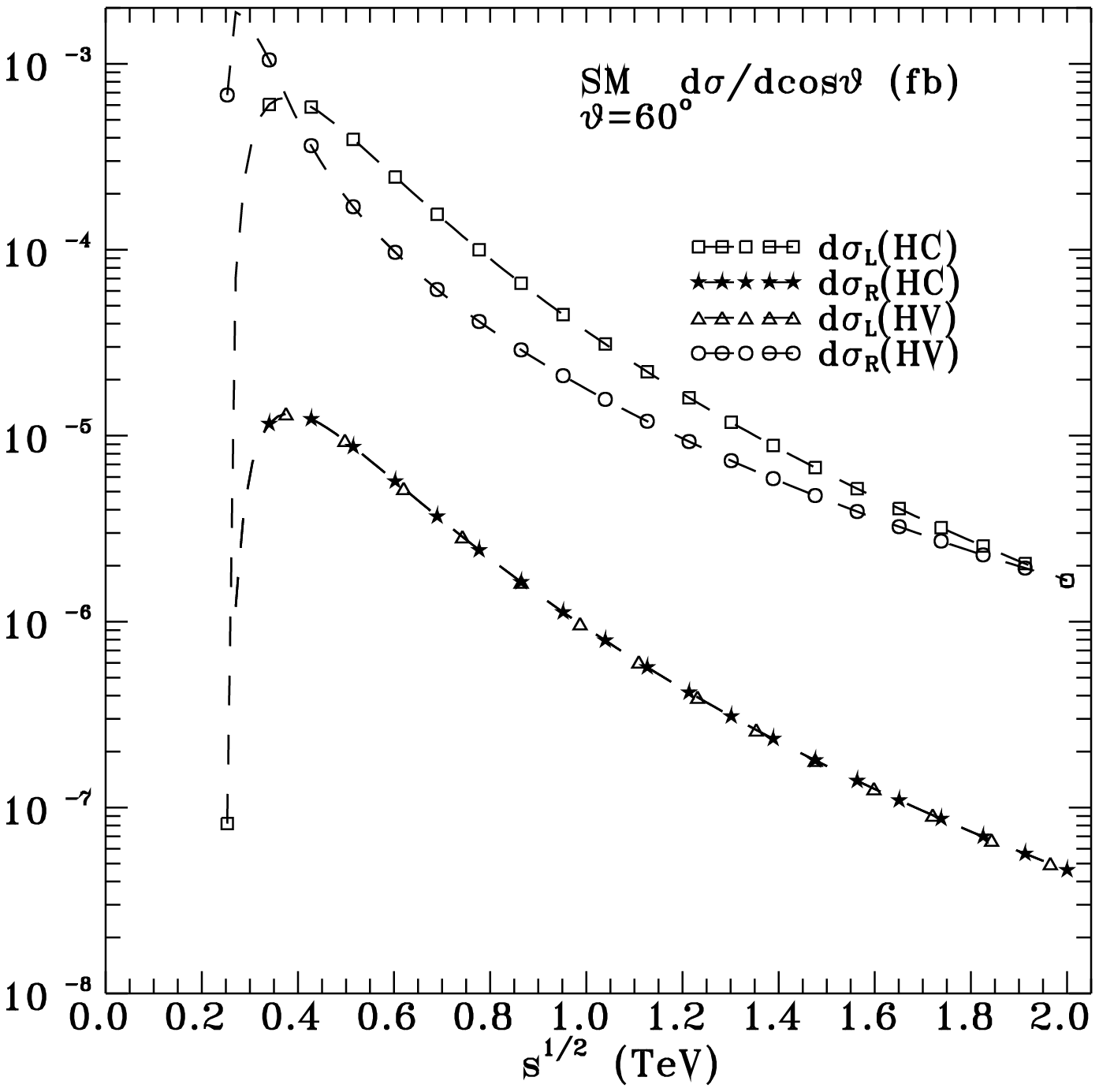, height=7.cm}\hspace{0.5cm}
\epsfig{file=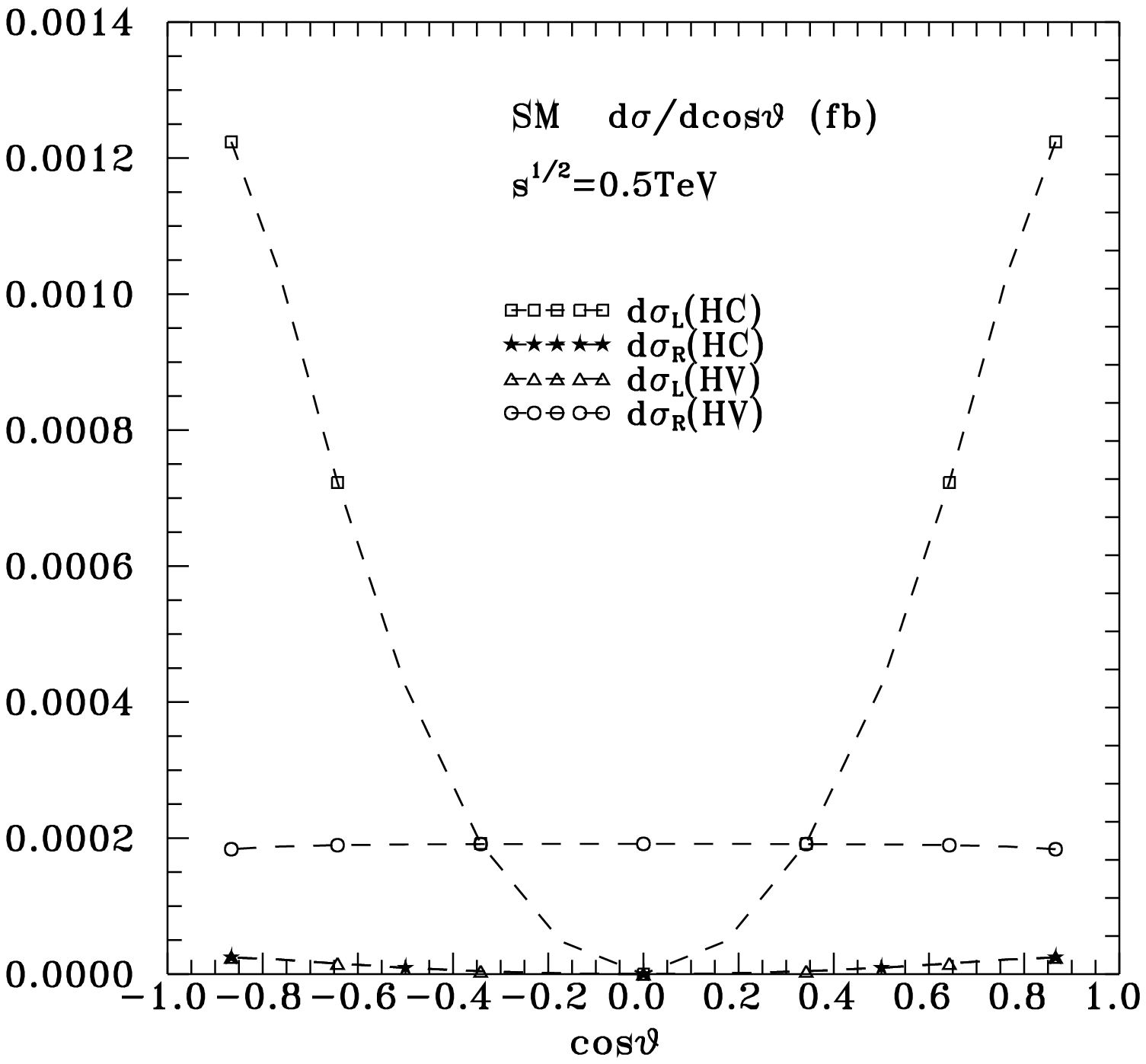,height=7.cm}
\]
\[
\epsfig{file=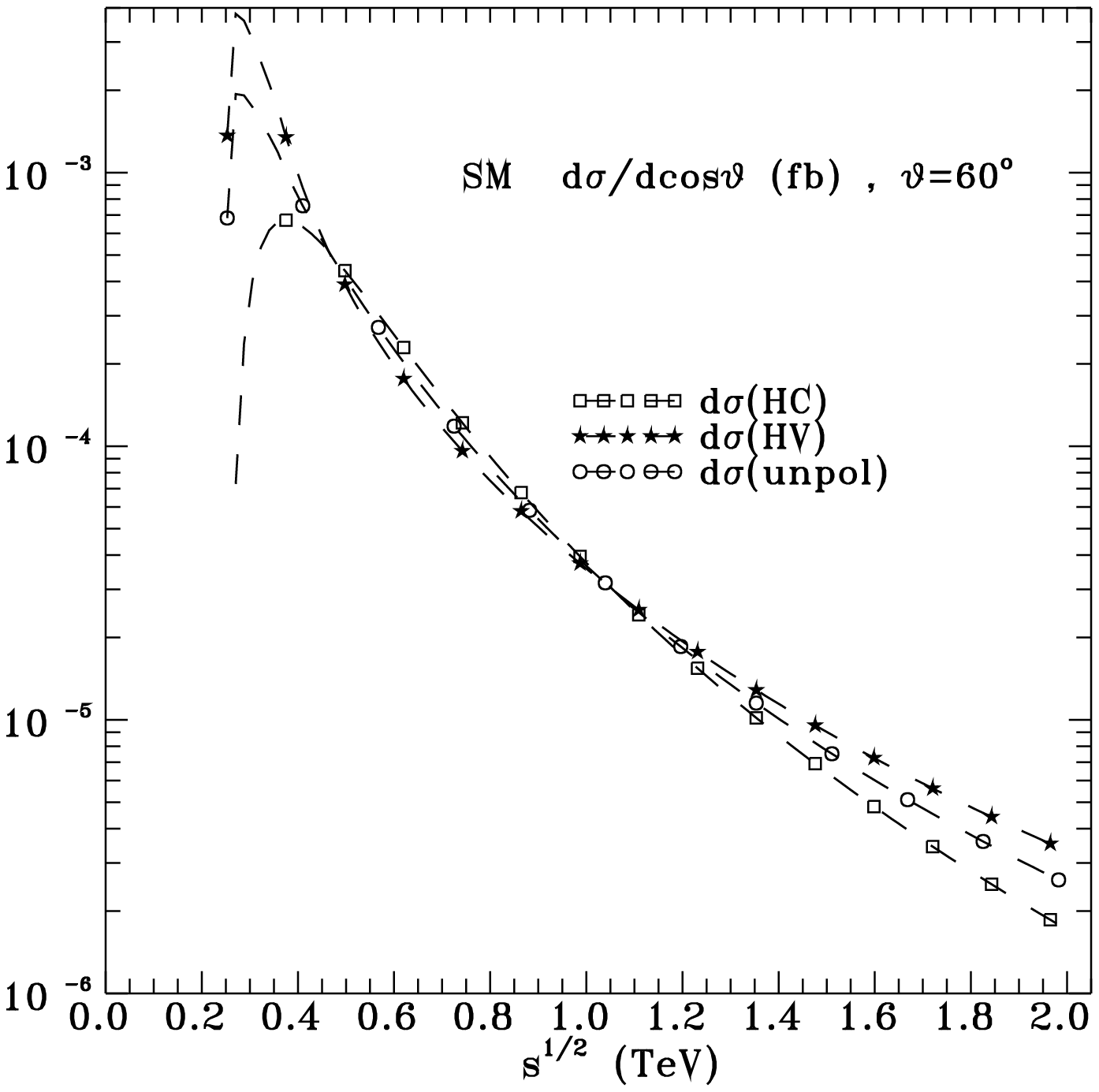, height=7.cm}\hspace{0.5cm}
\epsfig{file=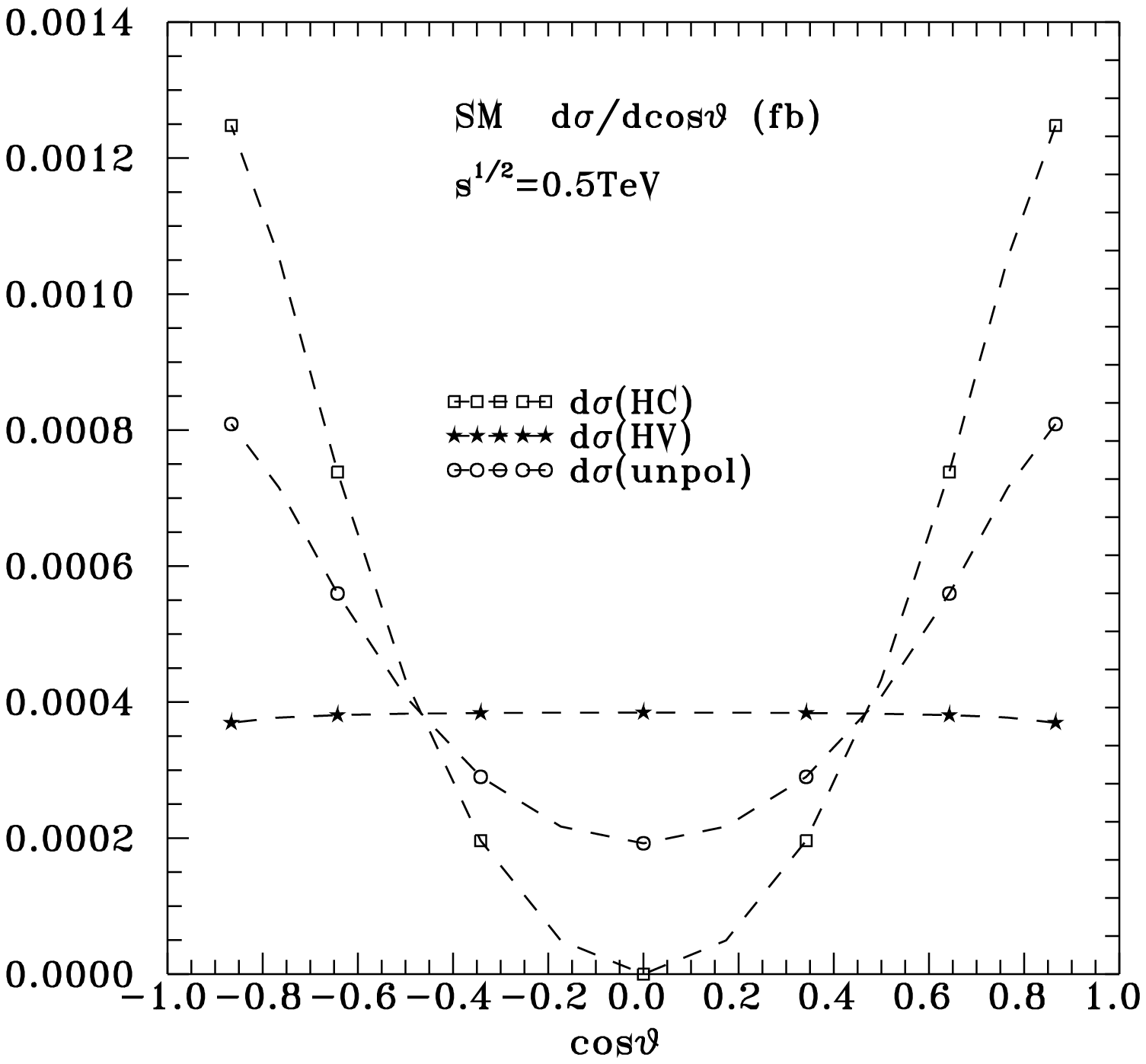,height=7.cm}
\]
\caption[1]{SM 1-loop differential cross sections  as defined in (\ref{HV-sigma}, \ref{HC-sigma}).
Left panels present the energy dependencies at $\theta=60^\circ$, while right panels the angular
ones at $\sqrt{s}=0.5 {\rm TeV}$.}
\label{Fig2}
\end{figure}

\clearpage

\begin{figure}[h]
\vspace{-1.2cm}
\[
\epsfig{file=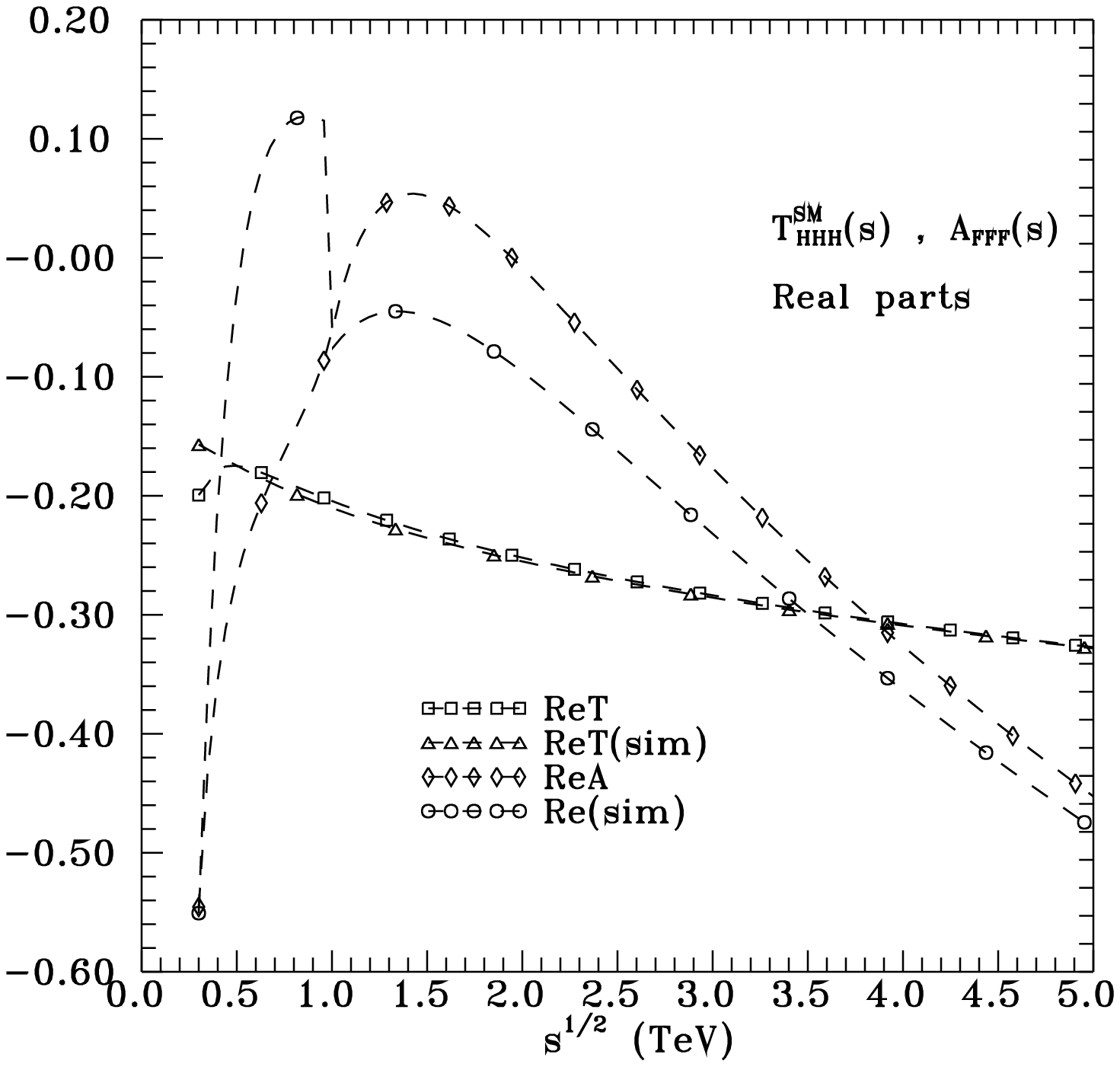, height=6.3cm}\hspace{0.5cm}
\epsfig{file=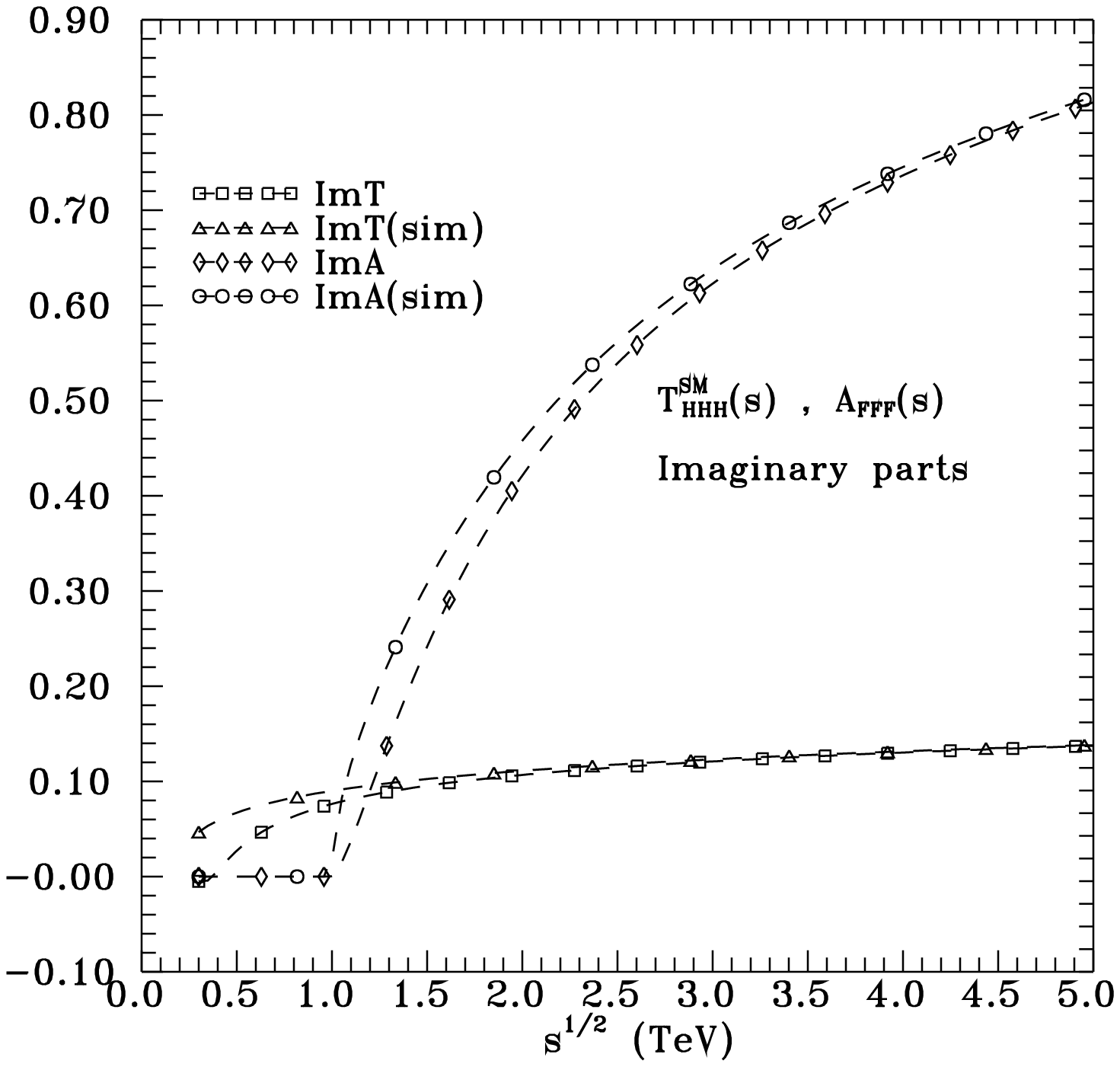, height=6.3cm}
\]
\[
\epsfig{file=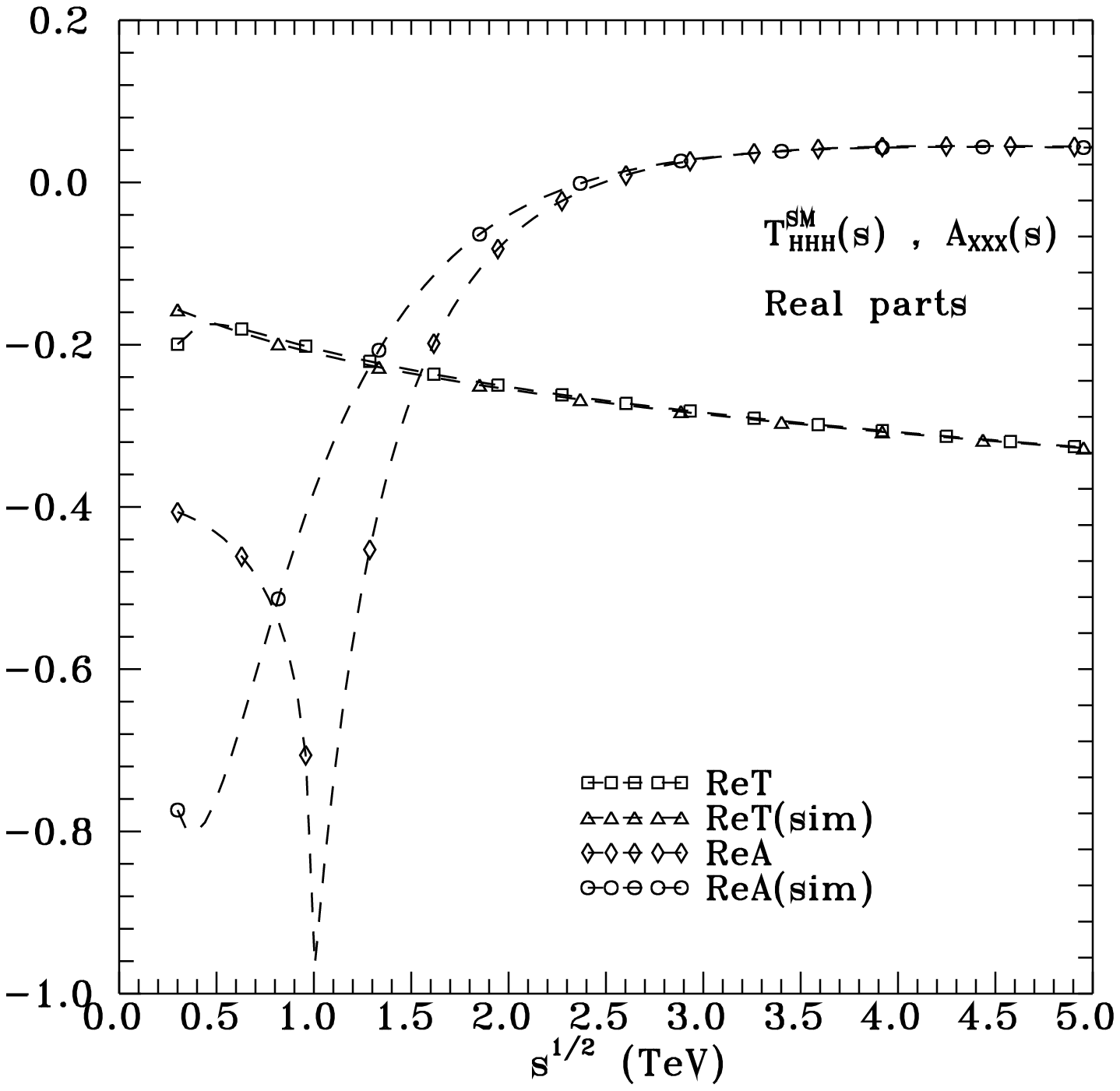, height=6.3cm}\hspace{0.5cm}
\epsfig{file=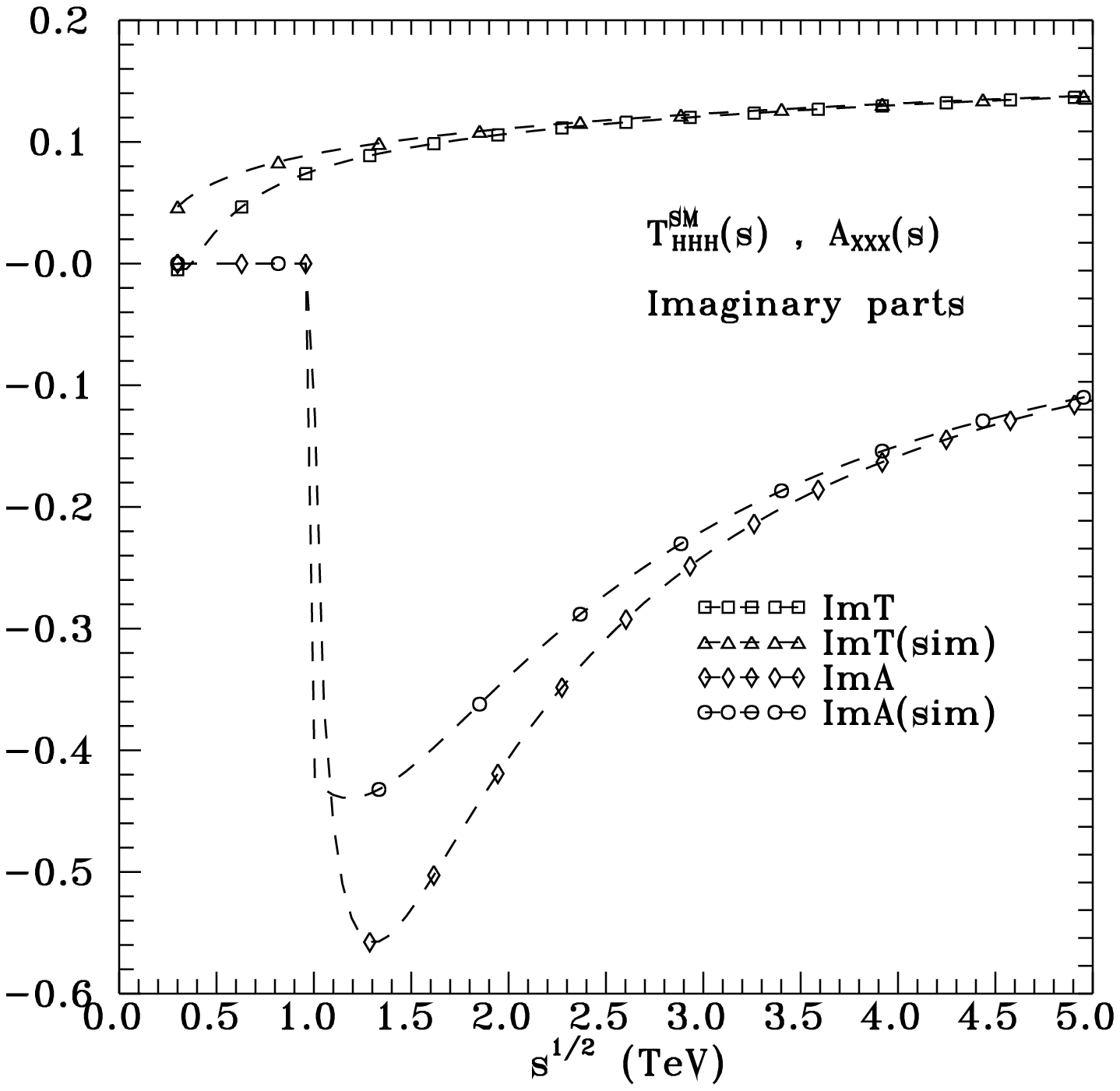, height=6.3cm}
\]
\[
\epsfig{file=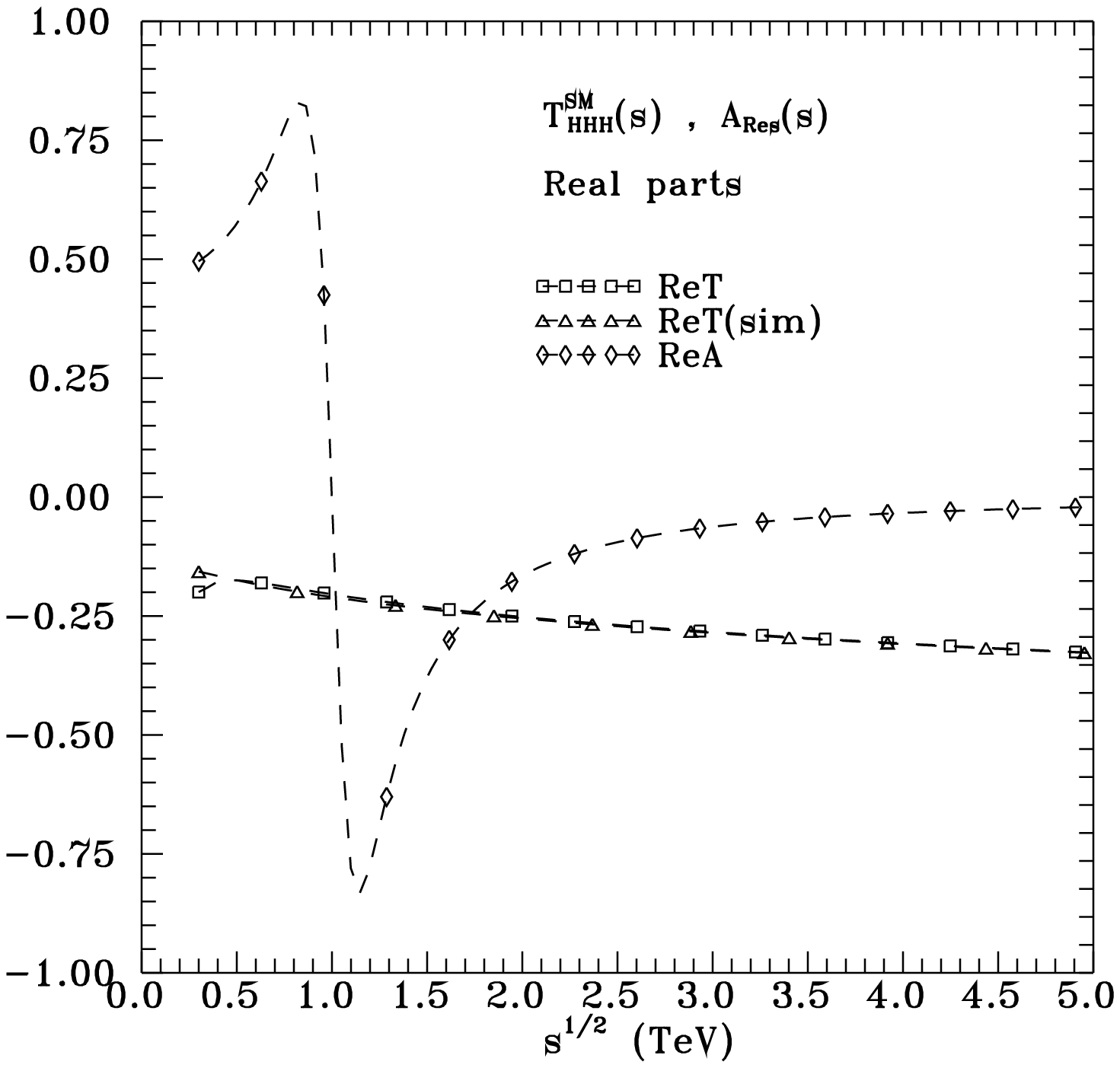, height=6.3cm}\hspace{0.5cm}
\epsfig{file=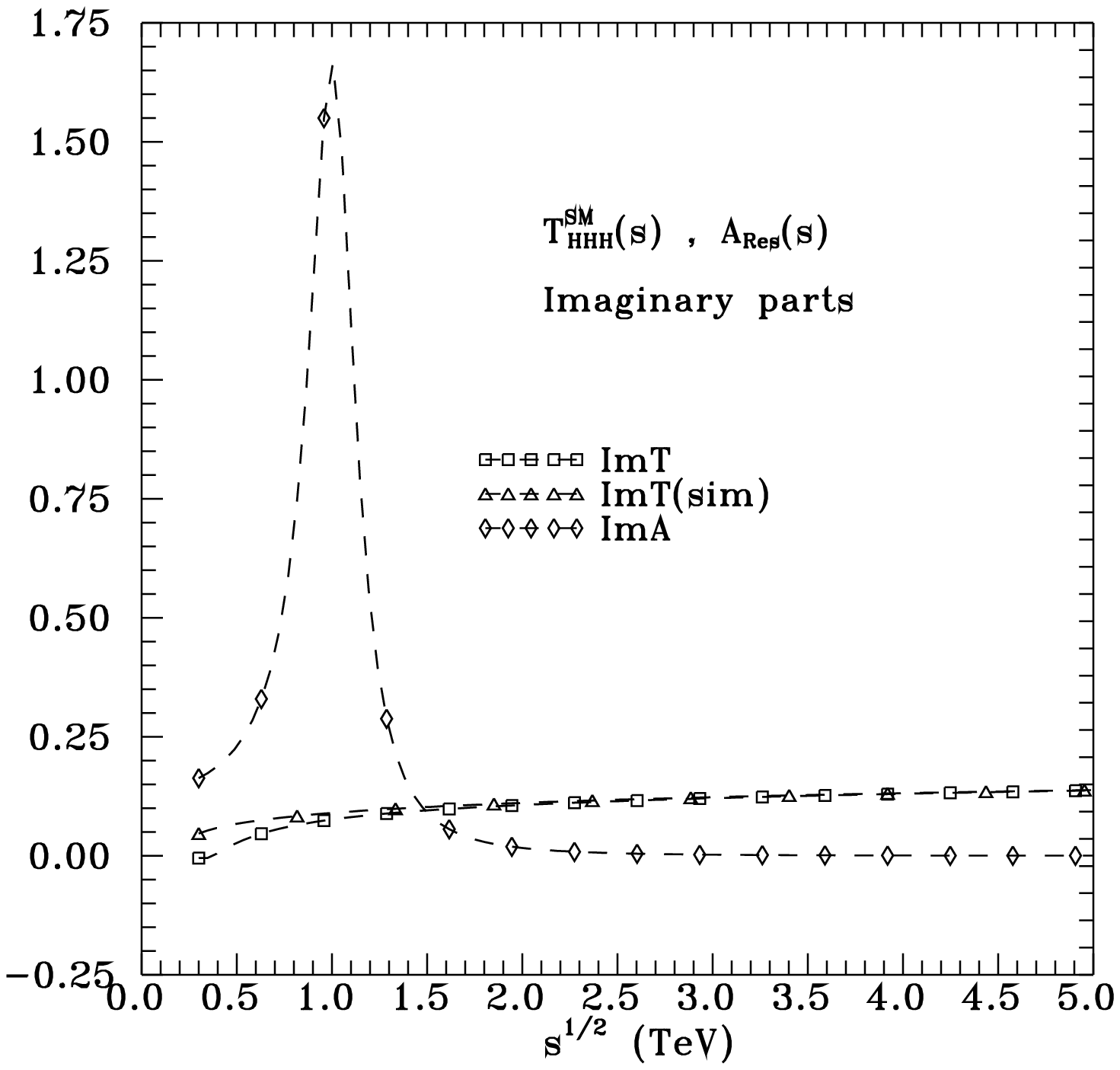, height=6.3cm}
\]
\caption[1]{The $s$-dependence  of SM form factor $T^{SM}_{HHH}$ defined in (\ref{T1loopHHH}),
together with the new physics contributions to it from  $A(s)_{FFF}$ of (\ref{ANPFFF}) (upper panels),
$A(s)_{XXX}$ of (\ref{ANPXXX}) (middle panels), and $A(s)_{\rm Res}$ of (\ref{ANPRes}) (lower panels).
Left and right panels present real and imaginary parts respectively. T refers to the SM contribution,
A to the new physics contributions, with the following parameters
$m_X=0.5$ TeV and $g_{HXX}=-10$ TeV,
$m_F=0.5$ TeV and $g_{HFF}=-4$,
$M_R=1$ TeV, $\Gamma_R=0.3$ TeV, $g_{HR}g_{RHH}=0.5$ TeV.
The supersimplicity (SRS) predictions given in (\ref{Sud-ln2}, \ref{Sud-ln}) are denoted T(sim).}
\label{Fig3}
\end{figure}

\begin{figure}[p]
\[
\epsfig{file=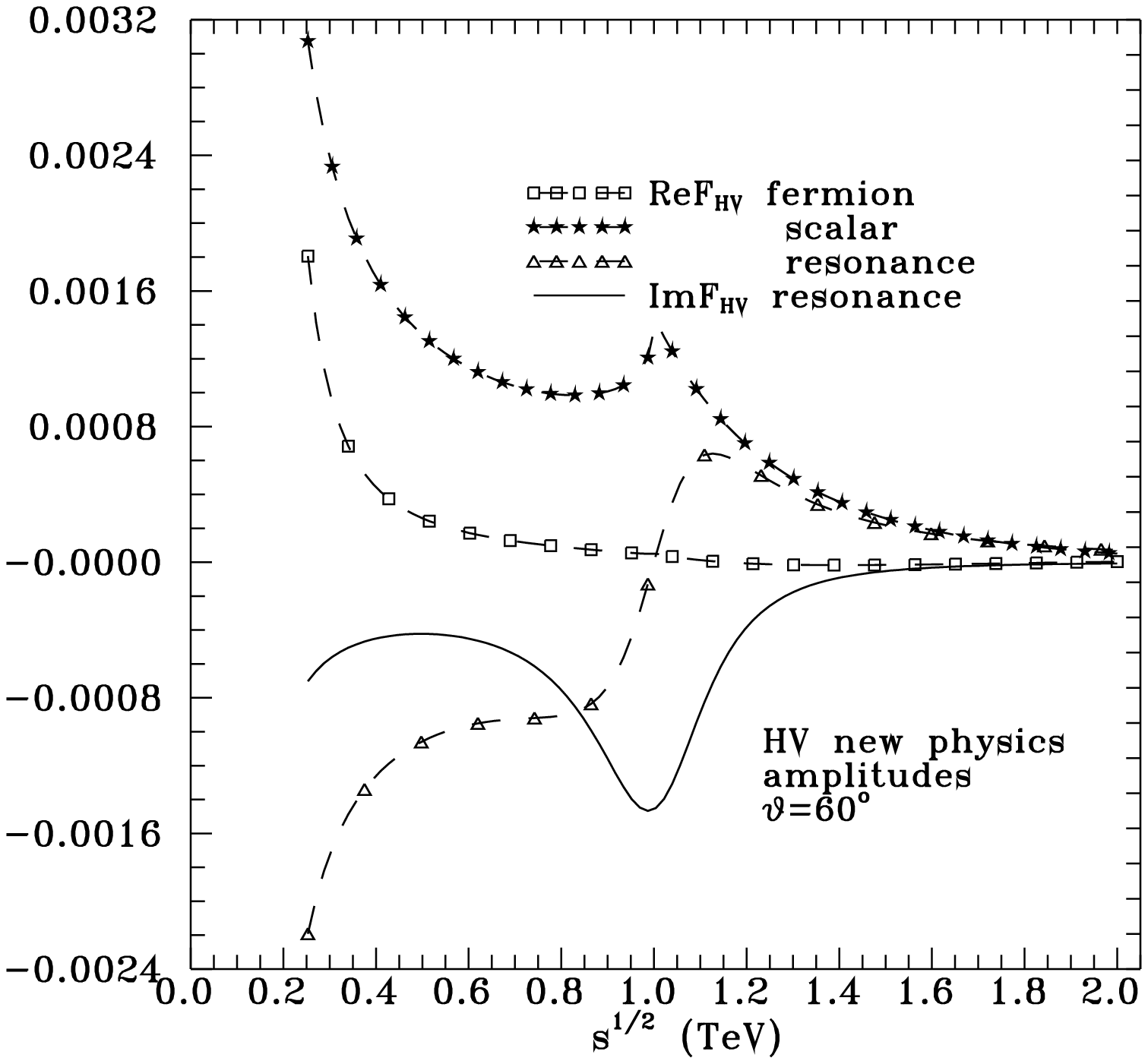, height=7.cm}\hspace{0.5cm}
\epsfig{file=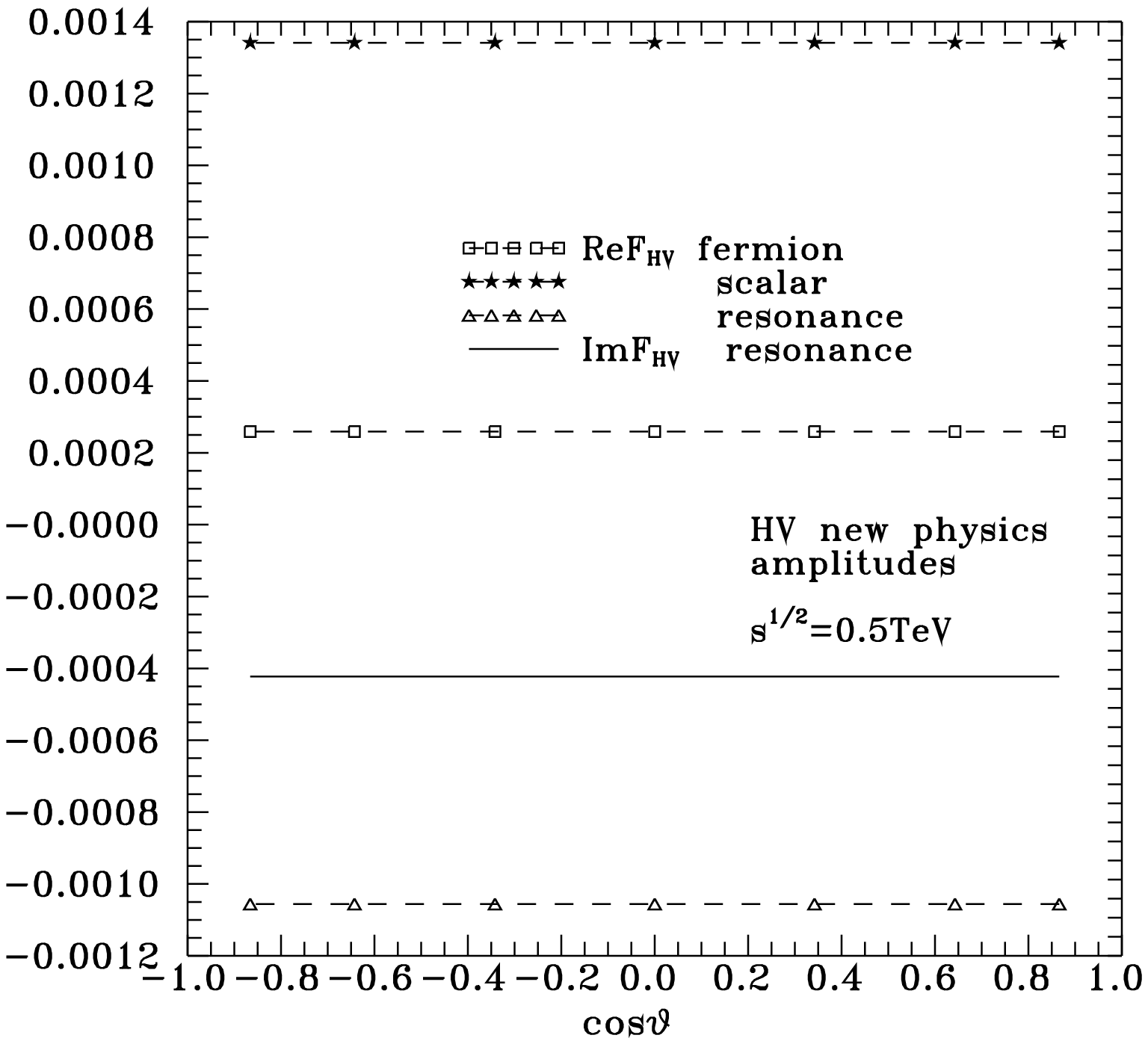, height=7.cm}
\]
\caption[1]{The New Physics contributions to the HV amplitudes induced by the HHH form factors
of Fig.3. The HC amplitudes and the  imaginary parts in the "new fermion" and "new scalar" HV amplitudes
 are vanishing and they are not shown. Left panel presents the energy dependencies, while the right
 panel the angular ones, as in Fig.1.}
\label{Fig4}
\end{figure}

\begin{figure}[p]
\vspace{-1.7cm}
\[
\epsfig{file=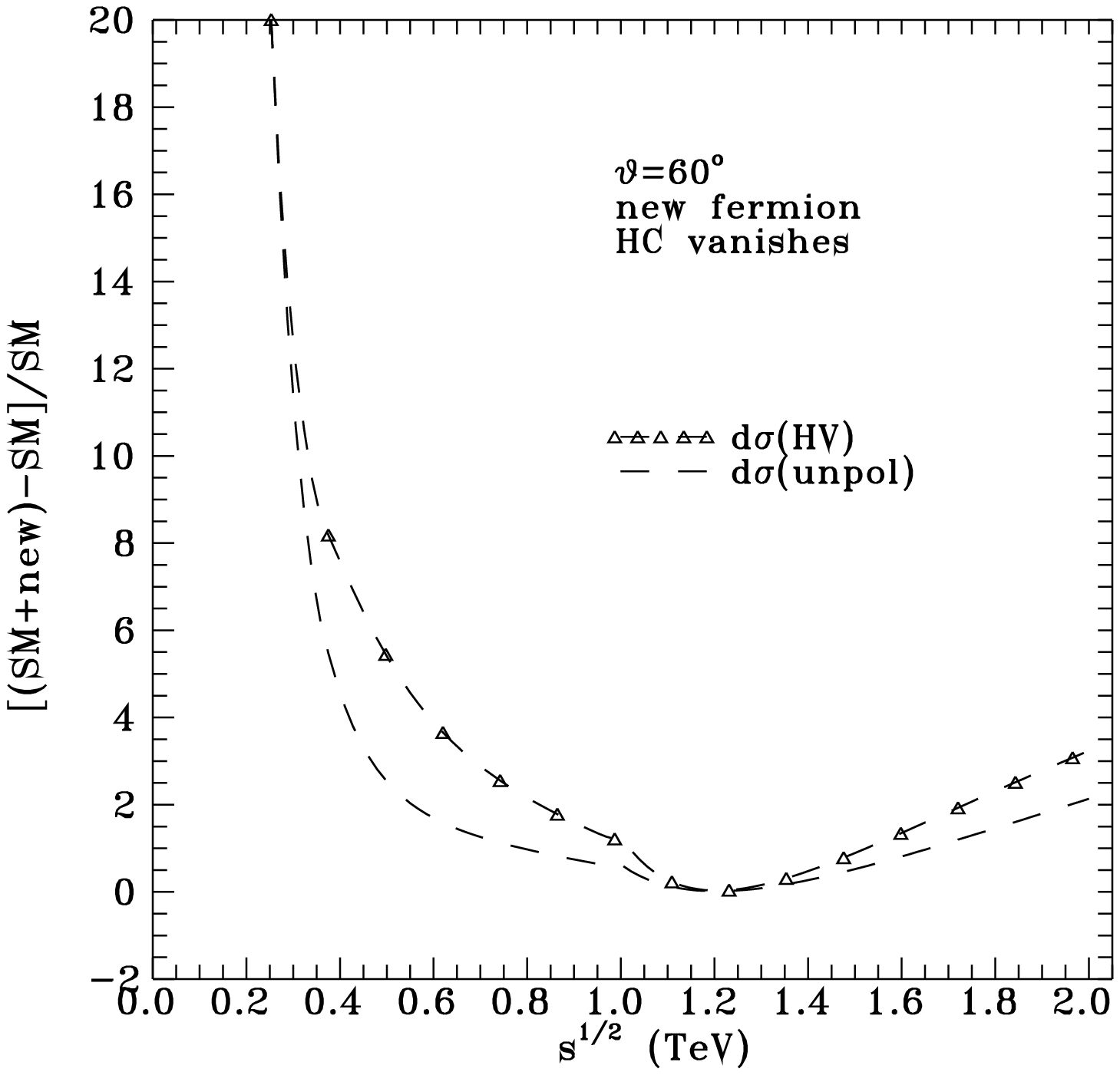, height=6.cm}\hspace{0.5cm}
\epsfig{file=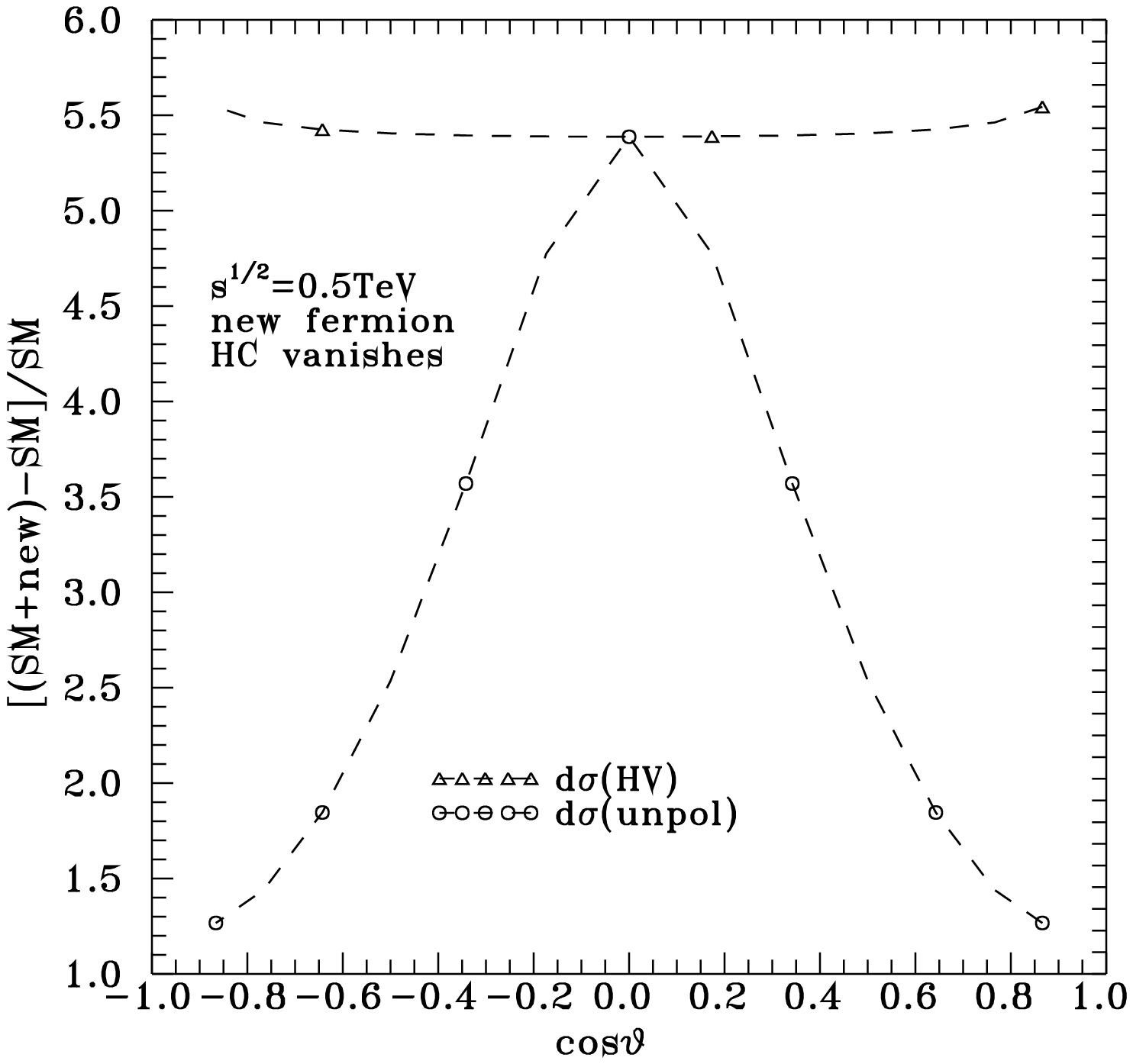,height=6.cm}
\]
\[
\epsfig{file=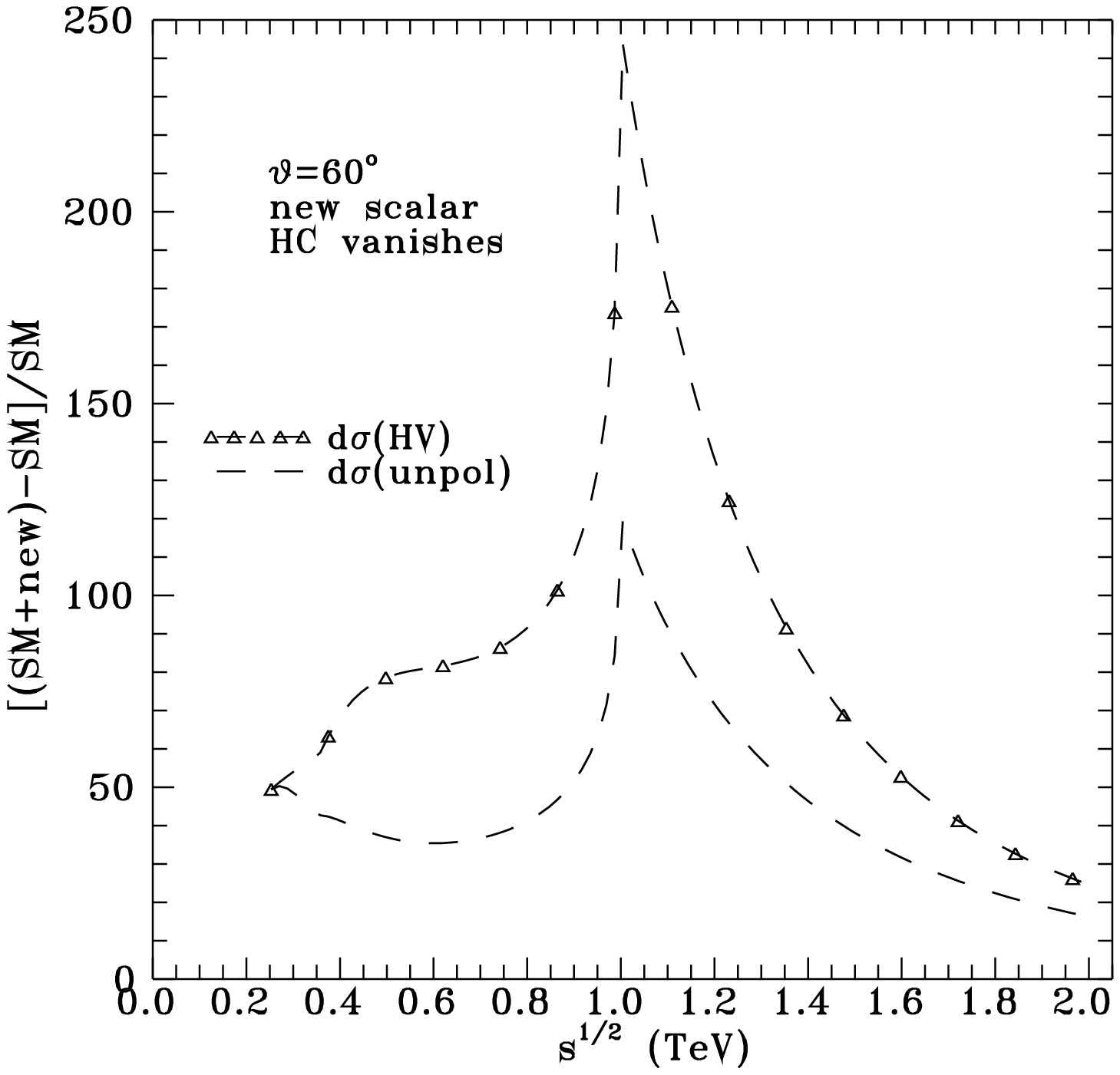, height=6.cm}\hspace{0.5cm}
\epsfig{file=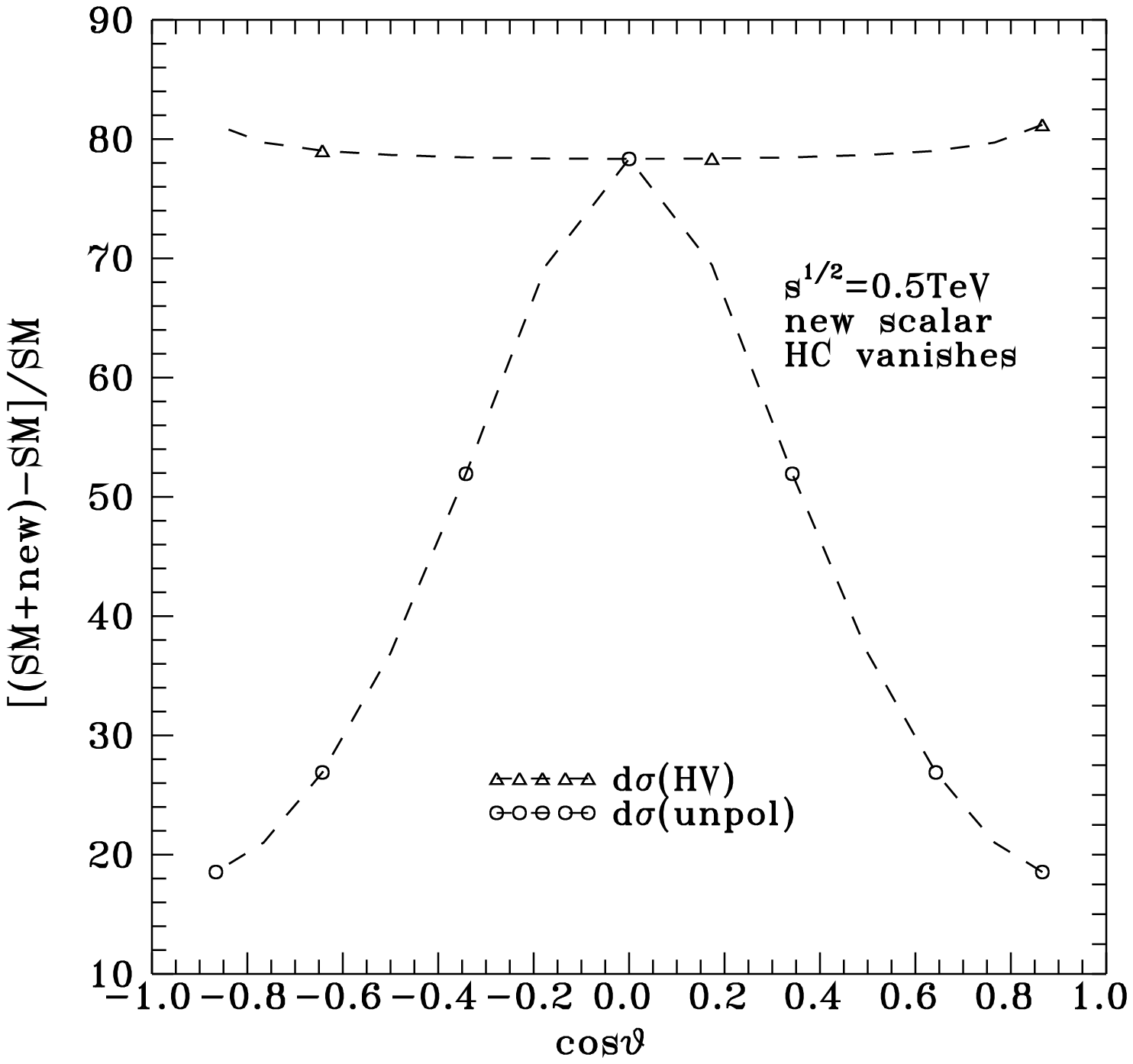,height=6.cm}
\]
\[
\epsfig{file=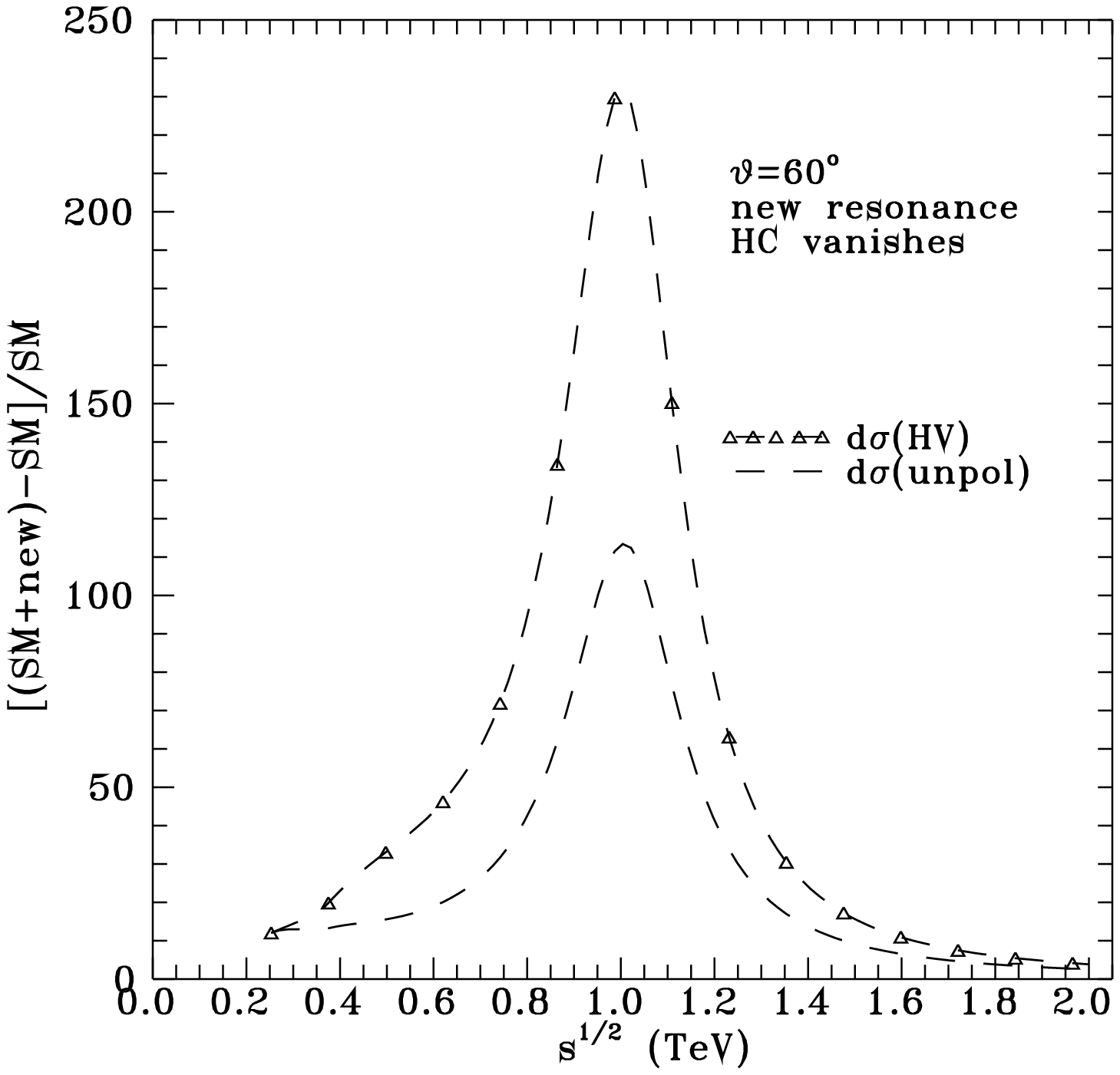, height=6.cm}\hspace{0.5cm}
\epsfig{file=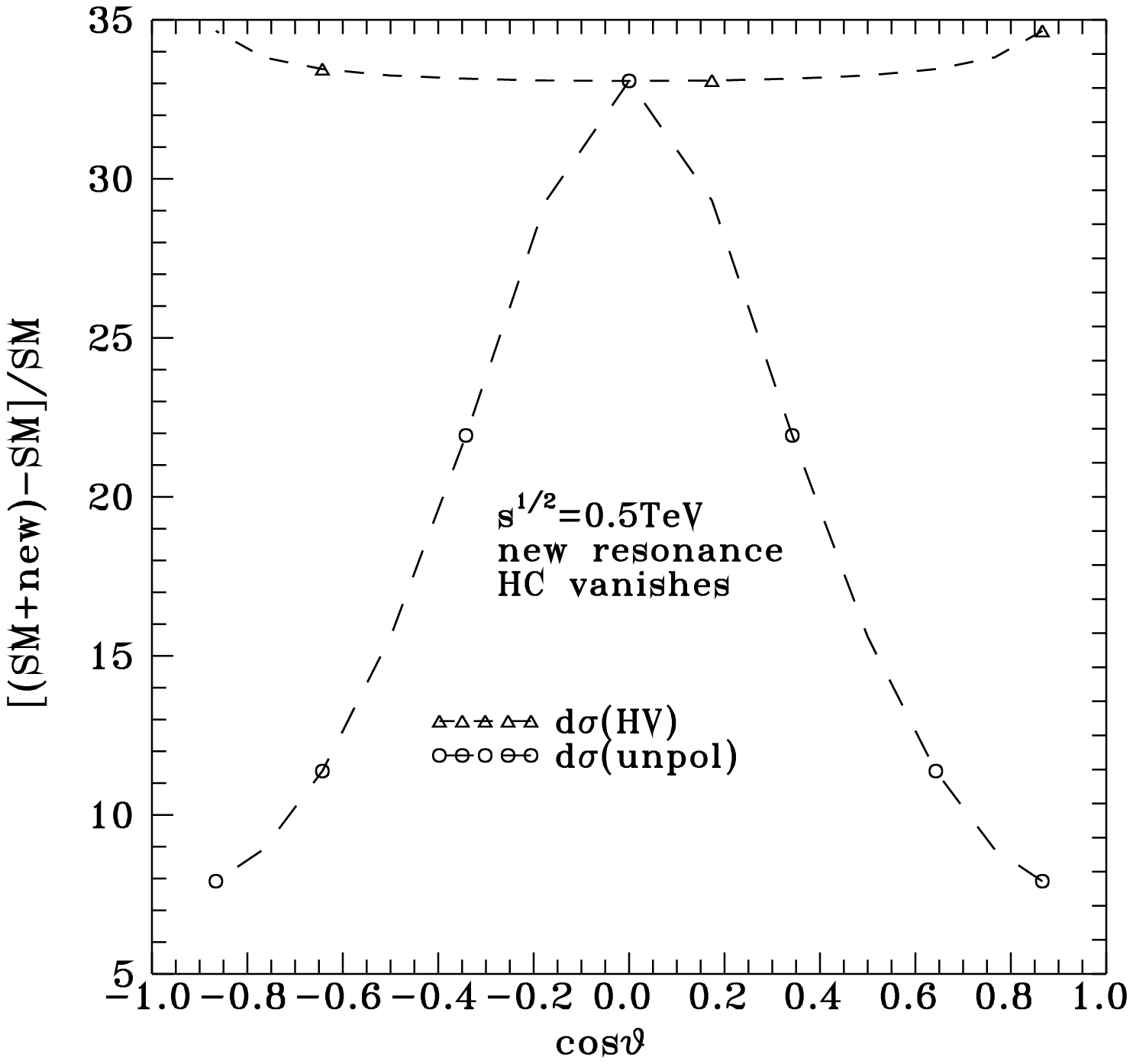,height=6.cm}
\]
\caption[1]{Dependencies of
$[d\sigma(SM+{\rm new})/d\cos\vartheta -d\sigma(SM)/d\cos\vartheta]/[d\sigma(SM)/d\cos\vartheta]$,
on the "new fermion" (upper), "new scalar" (middle) and the "new resonance" (lower panel)
contributions for the total HV case in (\ref{HV-sigma}) and for the unpolarized case.}
\label{Fig5}
\end{figure}

\end{document}